\documentclass[12pt]{article}

\usepackage{amsmath}

\begin{document}



\begin{center}
{\bf

\thispagestyle{empty}

BELARUS NATIONAL ACADEMY OF SCIENCES

B.I. STEPANOV's INSTITUTE OF PHYSICS

}
\end{center}

\vspace{50mm}
\begin{center}

{\bf

A.A. Bogush, V.V. Kisel, N.G. Tokarevskaya, V.M. Red'kov\footnote{E-mail: redkov@dragon.bas-net.by }\\
\vspace{5mm}

PETRAS's THEORY OF A SPIN-1/2 PARTICLE IN ELECTROMAGNETIC AND
GRAVITATIONAL FIELDS}

\vspace{5mm}

\end{center}

\begin{quotation}

20-component Petras' theory  of 1/2-spin  particle with  anomalous
magnetic momen\-tum  in presence of  external electromagnetic  and gravitational
fields is investigated. The gravitation field
is described as space-time curvature.
Correctness of the constructed equations in the sense  of general relativity
and gauge local Lorentz group symmetry is proved in detail.  Tetrad $P$-symmetry
of the equations is demonstrated.
A generally covariant representation of the invariant
bilinear form matrix is established and the  conserved current
of the 20-component field  is constructed.
It is shown that after exclusion  of the additional vector-bispinor
$\Psi_{\beta}(x)$  the wave equation for the principal $\Psi$-bispinor looks as
generally covariant Dirac's equation with electromagnetic minimal and Pauli
interactions and with an additional gravitational interaction through  scalar
curvarture $R(x)$-term. The massless  case is analyzed in detail.
The conformal non-invariance of the massless equation is demonstrated and new
conformally invariant equations for  20-component field are  proposed.

\end{quotation}

33 pages , references -- 80
\newpage

\begin{center}

{\bf CONTENT}

\end{center}

\noindent
\vspace{5mm}
Introduction

\vspace{5mm}
\noindent
1. Petras equation in flat space-time

\vspace{5mm}
\noindent
2.  Extension to general relativity

\vspace{5mm}
\noindent
3.
$SL(2.C)$ group and  gauge properties of spinor
connections

\vspace{5mm}
\noindent
4.  Invariant form matrix and conserved current

\vspace{5mm}
\noindent
5. Equation for a main bispinor $\Psi$

\vspace{5mm}
\noindent
6. Masless limit and conformal invariance

\vspace{5mm}
\noindent
7.  On gauge $P$-symmetry of the theory

\vspace{5mm}
\noindent
8.  On matrix formulation of the   20-component theory

\vspace{5mm}
\noindent
9.
Bilinear combinations in a Riemannian space-time

\vspace{5mm}
\noindent
10.
On canonical energy-momentum tensor

\vspace{5mm}
\noindent
11.
On energy-momentum tensor in spin-vector approach

\vspace{5mm}
\noindent
12.
Conclusion

\vspace{5mm}
\noindent
Supplement A. Ricci tensor and conformal transformation

\vspace{5mm}
\noindent
References

\newpage

\newpage
\subsection*{Introduction}

In order to have a comprehensive  theory of higher-spin fields, it
is necessary to be able to describe interactions. The most
important and the best understood is, of course,  the
electromagnetic. But the gravity is very important as well, at
least from theoretical viewpoint if not from practical. Generally,
a standpoint may be brought  forward that we should give much
attention to those facts of physics in the Minkowski space-time
which allow for  extension to a generally covariant physics,
because only doing so we will be able to reconcile  the principles
of particle physics with those of general relativity, and thus will be able to construct a deeper theory.

Else one point should be taken as of principal importance. Since  wave equations with subsidiary
conditions usually lead to consistency  difficulties
when  minimally coupled  to an electromagnetic  field,
it seems best to avoid subsidiary  conditions from the start. Furthermore, it seems the best to start with
a first-order system, which is  linked up to a certain Lagrange function, and in which some physically
required restrictive conditions  are built in from the very beginning in accordance with the  Pauli-Fierz
[1,2] approach.   The more so is in  the context of possible background of a non-Euclidean geometry,
in view of arising additional subtleties in a consistency problem.

It is well known that if one looks for a first-order  differential equation describing a mass $m$ spin $s$
field,  that is form-invariant  under Lorentz transformations,  derivable from a Lagrangian, then  one  does not
uniquely obtain the common Dirac or Duffin-Kemmer or some other equations.
We have taken the existence of such generalized particles for granted, relying for justification
on our experience  with Dirac and electromagnetic fields. It does
not seem to us to appeal always  to experiment; after all, purpose in theoretical physics is not
to describe the world as we find it, but to explain -- in terms of
a few fundamental principles --  why the world is just the way it is.

A general theory of such first-order and Lagrangian-based equations has been treated at great length by
Fierz and Pauli [1,2], Dirac [3], Bhabha [4], Harish-Chandra [5] and many others  [6-12].
In fact, it was shown that almost an infinite many  of such equations is possible. However, quite
recently there had been a notable dearth of examples of such theories that had been developed
at a large extent comparable  to the Dirac or Duffin-Kemmer examples\footnote{
In recent years some interest
in the Duffin-Kemmer-Petiau formalism again can be noted -- see [40-45].}.
But  such  particle models, being elaborated in full  detail, might shed  new light on the theory of general
arbitrary-spin wave equations.

Much work in this direction has been done [13-39]. For instance, generalized
wave equations for  particles of spin 1/2 (Petras, Halhil, Capri, Fedorov, Pletjuchov, Bogush,
Kisel [15,17,21,22,24,28,30,34,38,39]
 with arbitrary anomalous magnetic momentum
have been worked out.
 Else one extension, less known, has been done for a boson particle:
a next simplest theory beyond the Duffin-Kemmer case (of spin 0 and 1) was created by Fedorov and Pletjuchov
[25,26] and an explicit representation of the basic first-order equation was given.

In this paper we examine the properties of the 20-component  spin 1/2 theory  in the case of no
interaction as well as  in presence of external electromagnetic and gravitational fields.
Why another  paper on this matter. The reader  can choose from a number of them.
Another work will be worth while only  if it offers something new in content and perspective.
It is known
that in  Minkowski space-time this theory  is very simply related to the Dirac equation. In the free-field case,  the two theory are
equivalent, and the more complicated 20-component model can be reduced to the ordinary 4-component one.
In the presence of external electromagnetic fields, one  obtains  a Dirac-Pauli theory with an
additional term corresponding to an anomalous magnetic momentum\footnote{This else one
time illustrates explicitly a point emphasized by Wightman that although two field theories  may be
equivalent in absence of interaction, they may be completely inequivalent in  presence of
interaction.}. So, starting from the principles about which we are most certain, relativity, quantum mechanics,
and the first order equation theory, an additional interaction term, introduced in quite
phenomenological way, emerges anew from prime principles  as a natural consequence.
It turns out that consistent treatment of the 20-component  fermion
theory lead us to a Dirac-Pauli fermion which interacts
with gravitational field in a manner different to an ordinary
Dirac's particle: through an additional Ricci scalar term. So, the
rationale for additional Pauli term might  be, reasoning in  somewhat speculate manner,
found in consistent description of additional gravitational interaction of the fermion
through a Ricci scalar.

The present article aims to provide  a self-contained, comprehensive, an  up-to-date exposition of the matter.
So, the development is detailed and logical throughout, with  each step carefully motivated by what has gone
before, and emphasizing the reason why such a step should be done.
As to  content, although a paper contains a good amount of new material, the taking into account
of the possible presence of gravitational background is  its most distinctive thing.
The more is so, because the formalism developed may potentially
find applications far removed from its original scene of the use.
This in part explains that frequently we immersed ourselves in technicalities of specific problems;
and we took a firm hold on this line. So we have tried to err on the side of inclusion rather than
exclusion\footnote{We hope this work will suit those who are ready to look with interest at old and
conventional methods of work with minimum of modern and rigorous mathematics, in our opinion much of those
should be revived.}.
We hope that we have improved on the original literature
in several places, as for instance, in Lorentz group-based techniques used in general covariant description
of a fermion field; analysis of the energy-momentum tensor of the generalized fermion in a curved space-time;
in deriving a generally covariant Bargmann-Mishel-Telegdi equation for a fermion with anomalous
magnetic moment (this part of the work will be done later and separately); in separating the variables in generally covariant wave equations, and so on.
We have tried give citations on topics that are mentioned here.
But we did not always know who was responsible  for material presented here, and the mere absence
of a citation should not be taken as  a claim that material presented here is original.
But hoping  some of it is.

\subsection*{1. Petras equation in flat space-time}

Original 20-component equation for a  spin 1/2 particle is taken in the form
$$
\gamma^{a} \partial_{a} \Psi + \mu \; (K g^{ab} + A \sigma^{ab}) \; \partial_{a} \Psi_{b}  = M \Psi  \; ,
\eqno(1.1a)
$$
$$
\mu \; ( N \delta^{c}_{b} + B \sigma_{b}^{\;\;c} ) \; \partial_{c} \Psi  = M \Psi_{b} \; .
\eqno(1.1b)
$$

\noindent
Here, a full wave function includes bispinor $\Psi (x)$ and vector-bispinor  $\Psi_{a}(x)$;
$\gamma^{a}$ denotes Dirac's  four-by-four matrices\footnote{Working in the Minkowski
space-time, the metric tensor $g^{ab} =diag (+1,-1,-1,-1) \; $ is used throughout. Latin letters
take the values  $0,1,2,3$, Greek letters are reserved for generally covariant indices.};
$\sigma^{ab} = {1 \over 4} (\gamma^{a} \gamma^{b}- \gamma^{b} \gamma^{a} )$; $mc/\hbar = iM$ ;
$\mu$ stands for an additional dimensionless characteristic of the fermion under consideration;
$K,N,A,B$ are certain real numbers.

Rewriting eq. (1.1b) as
$$
\Psi_{b}(x)= {\mu \over M } \; ( N \delta^{c}_{b} +
B \sigma_{b}^{\;\;c} ) \; \partial_{c} \Psi  \; ,
\eqno(1.2)
$$

\noindent one can exclude the vector-bispinor from eq. (1.1a):
$$
[ \;\gamma^{a} \partial_{a}  +  \mu^{2}  M^{-1}
\;(K g^{ab} + A \sigma^{ab})\; ( N \delta^{c}_{b} + B \sigma_{b}^{\;\;c} )\;
\partial_{a}  \partial_{c}   - M \; ]\; \Psi  = 0 \; .
\eqno(1.3)
$$

\noindent Introducing the notation
$$
Z^{ac} = (K g^{ab} + A \sigma^{ab})\; ( N \delta^{c}_{b} + B \sigma_{b}^{\;\;c} ) \; ,
$$

\noindent eq. 1.3) reads as
$$
[ \; \gamma^{a} \partial_{a}  + { \mu^{2} \over M} \;
Z^{ac} \; \partial_{a}  \partial_{c} - M \;] \; \Psi (x) = 0  \; .
\eqno(1.4)
$$

\noindent
Let us consider more closely the expression for $Z^{ac}$
$$
Z^{ac} = KN g^{ac} + (KB + NA) \sigma^{ac} + AB \; \sigma^{ab} \sigma_{b}^{\;\;c} \;,
$$

\noindent
on allowing for equalities
$$
\sigma^{ab} \sigma_{b}^{\;\;c} =
( {1 \over 2} \gamma^{a} \gamma^{c} + {1 \over 4} g^{ac}) \; , \qquad
g^{ac} = {1 \over 2} ( \gamma^{a} \gamma^{b} + \gamma^{b} \gamma^{c} ) \; ,
$$

\noindent the $Z^{ac}$  takes on the form
$$
Z^{ac} = (KB + NA) \; \sigma^{ac} \;  +
$$
$$
+ \gamma^{a} \gamma^{c} \;  ( {1 \over 2} KN + {1 \over 2} AB + {1 \over 8} AB ) \; +
\gamma^{c} \gamma^{a} \;  ( {1 \over 2} KN + {1 \over 8} AB ) \; .
$$

\noindent Numerical parameters are to be chosen that the $Z^{ac}$
coincides, apart from a numerical factor, with the skew combination $\sigma^{ac}$.
This will be so if the following requirements hold
$$
 ( {1 \over 2} KN + {5 \over 2} AB  ) = + C \; , \qquad ( {1 \over 2} KN + {1 \over 8} AB ) = -C \; ,
\eqno(1.5a)
$$
\noindent then
$$
Z^{ac} = [\; (KB + NA) + 4C\; ] \; \sigma^{ac} \; .
\eqno(1.5b)
$$

\noindent From  eq. (1.5a) it follows
$$
4KN + 3AB = 0 \; , \; C = - { KN \over 3} \; .
\eqno(1.5c)
$$

\noindent Petras  [15] choose a solution in the form
$$
K = \sqrt{3} \; , \; N = \sqrt{3} \; , \; A = +2 \;, \; B = - 2 \; , \;
Z^{ac} = -4 \sigma^{ac} \; .
\eqno(1.6)
$$

\noindent Correspondingly, eqs.  (1.1) will read as
$$
\gamma^{a} \partial_{a} \Psi + \mu \; (\sqrt{3} \; g^{ab} +
2 \sigma^{ab}) \;  \partial_{a} \Psi_{b}  = M  \; \Psi  \; ,
\eqno(1.7a)
$$
$$
\mu \; ( \sqrt{3} \; \delta^{c}_{b} - 2 \sigma_{b}^{\;\;c} ) \;  \partial_{c} \Psi  = M  \; \Psi_{b} \; .
\eqno(1.7b)
$$

\noindent These are equivalent to the following
$$
\Psi_{b} = { \mu \over M} \;
( \sqrt{3} \; \delta^{c}_{b} - 2 \sigma_{b}^{\;\;c} ) \; \partial_{c} \Psi  \; ,
\eqno(1.8a)
$$
$$
[\; \gamma^{a} \partial_{a}  - {4 \mu^{2} \over M} \;
\sigma^{ac} \; \partial_{a} \partial_{c} - M  \;] \;\Psi  = 0 \; .
\eqno(1.8b)
$$

In presence of an external electromagnetic field one is to lengthen derivatives
(the combination  ${ e \over \hbar c} $ will be denoted as  как $g$)
$$
\partial_{a} \Longrightarrow D_{a} = \partial_{a} - i {e \over \hbar c}  A_{a}\; .
$$

\noindent at this eqs.  (1.7) transform into
$$
\gamma^{a} D_{a} \Psi + \mu \; (\sqrt{3} g^{ab} +
2 \sigma^{ab})\; D_{a} \Psi_{b}  = M \Psi  \; ,
\eqno(1.9a)
$$
$$
\mu \; ( \sqrt{3} \delta^{c}_{b} - 2 \sigma_{b}^{\;\;c} ) \; D_{c} \Psi  =
M \Psi_{b} \; .
\eqno(1.9b)
$$

\noindent Obviously, eqs. (1.9)  are equivalent to
$$
\Psi_{b} = { \mu \over M} \;
( \sqrt{3} \delta^{c}_{b} - 2 \sigma_{b}^{\;\;c} ) D_{c} \Psi  \; ,
\eqno(1.10a)
$$
$$
[ \; \gamma^{a} D_{a}  - {4 \mu^{2} \over M} \;
\sigma^{ac} D_{a} D_{c}  - M ] \; \Psi  = 0 \; .
\eqno(1.10b)
$$

\noindent
Since the lengthened derivatives do not commute mutually, a relation
$
4 \sigma^{ac} D_{a} D_{c} = - 2 ig \sigma^{ac} F_{ac} \; , \;
$
holds, where  $F_{ac} = \partial_{a} A_{b} - \partial_{b} A_{a}$ stands for the electromagnetic tensor.
Correspondingly, eq.  (1.10b) will take on the Dirac-Pauli  form
$$
 (\; \gamma^{a} D_{a}  + i  {2 g \mu^{2} \over M} \;
\sigma^{ac} F_{ac} -  M  \; ) \; \Psi (x)  = 0 \; ,
\eqno(1.11)
$$

\noindent with an additional interaction term generated by an anomalous magnetic momentum.
In dimension unites eq. (1.11) reads
$$
 (\; i \gamma^{a} D_{a}  - i  {2 e \mu^{2} \over  mc^{2} } \;
\sigma^{ac} F_{ac}   -   {mc\over \hbar} \;  ) \; \Psi (x)= 0 \; .
\eqno(1.12)
$$

\noindent Let us look more closely at the matters of dimensions in eq. 91.12). As  (notation $[...]$
stands for 'dimension of (...)')
$$
[ { e \over \hbar c} A ] = {1 \over l } \Longrightarrow
[{e \over \hbar c } F ] = {1 \over l^{2} }  \Longrightarrow
{ [ e F ] \over  [m c^{2}]}  = {1 \over l} \;,
$$

\noindent the quantity  $\mu$ does   carry no physical dimension.

\subsection*{2.  Extension to  general relativity }

Now we are to proceed to the study the main question: given a first order  equation that is form invariant
under special relativity, what will it look like on the background of general relativity.

A first step, quite evident one in view of the known absolute necessity to use the vierbein formalism
of Tetrode-Weyl-Fock-Ivanenko  [46-80] at describing spinor fields on a curved space-time background,
is to introduce a new set of field functions:
$$
\Psi (x) , \; \Psi_{b}(x) \qquad \Longrightarrow \qquad  \Psi (x) ,\; \Psi_{\beta}(x) \; .
$$

\noindent At this,  $\Psi (s)$ transforms as scalar under general coordinate changes, and as
bispinor under
local Lorentz changes of local tetrads\footnote{Terms {\em tetrad} and {\em vierbein} will be used
interchangeably.}; in turn,  $\Psi_{\beta}(x)$ behaves as  a general covariant vector and as a local
tetrad bispinor at the same time.

A second step  is the postulating of a general covariant equation:
$$
\gamma^{\alpha} D_{\alpha} \Psi  + \mu
\; (\; \sqrt{3} g^{\alpha \beta}  + 2 \sigma^{\alpha \beta}  \;) \;
          D_{\alpha} \Psi _{\beta}  = M \Psi  \; ,
\eqno(2.1)
$$
$$
\mu \; ( \sqrt{3} \delta_{\beta}^{\;\;\rho} -
 2 \sigma_{\beta}^{\;\;\rho} ) \;  D_{\rho} \Psi  = M \; \Psi_{\beta}  \; .
\eqno(2.2)
$$

\noindent  Here,  $g^{\alpha \beta}(x) $ designates the metric tensor of
a curved space-time background; its vierbein make-up is as usual
$$
g^{\alpha \beta}(x)  = e^{\alpha} _{(a)}(x) e^{\beta}_{(b)}(x) \;  g^{ab} \; , \;
$$

\noindent $e_{(a)}^{\alpha}(x) $ stands for a tetrad. Generalized Dirac matrices are defined the relations
$$
\gamma^{\alpha}(x) = \gamma^{a} \; e_{(a)}^{\alpha} \; ,  \;
$$
$$
\sigma^{\alpha \beta} (x)  =
\sigma^{ab} \; e_{(a)} ^{\alpha}  e_{(b)}^{\beta}=
{1 \over 4} \; ( \gamma^{\alpha} \gamma^{\beta}
- \gamma^{\beta} \gamma^{\alpha}) \; .
$$

\noindent The symbol $D_{\alpha}$ stands for an extended covariant derivative
(acting differently on  bispinor and  vector-bispinor fields)
$$
D_{\alpha} = \nabla_{\alpha} + B_{\alpha}  -i g A_{\alpha} (x) \; .
$$

\noindent As usual, $\nabla_{\alpha}$ denotes the general covariant derivative;
$B_{\alpha}$ is called a bispinor connection
$$
B_{\alpha} = {1 \over 2} \; \sigma^{ab} \; e^{\beta}_{(a)} \;
\nabla_{\alpha} e_{(b) \alpha} = \gamma^{\alpha} \nabla_{\beta} \gamma_{\alpha} \; ;
$$

\noindent $A_{\alpha}$ represents an electromagnetic  4-vector.

To investigate symmetry properties of the equations  (2.1) and (2.2) it will be convenient  to exploit
the  known  spinor (Weyl's) basis  in the bispinor space defined by\footnote{Spinor indices
 $1,2, \dot{1}, \dot{2}$ will be suppressed in the following.}
$$
\gamma ^{a} = \left ( \begin{array}{cc}
                 0           &  \bar{\sigma}^{a}   \\
                 \sigma ^{a} &         0
                                        \end{array} \right ) \; ,
\sigma ^{a} = (I,\; +\sigma ^{k}) \; ,\;  \bar{\sigma }^{a} = (I, \;-\sigma ^{k} ) \; , \;
$$
$$
\Psi  =  \left ( \begin{array}{c}
               \xi    \\ \eta
                                \end{array} \right ) \;  ,  \;
\Psi_{\beta}  =
  \left ( \begin{array}{c}
               \xi_{\beta}    \\ \eta _{\beta}
                                         \end{array} \right ) \;  ,  \;
$$
$$
\xi  = \left ( \begin{array}{c}
              \xi ^{1}  \\ \xi ^{2}
                                   \end{array} \right ) \; , \;
\eta = \left ( \begin{array}{c}
          \eta _{\dot{1}} \\ \eta _{\dot{2}}
                                             \end{array} \right )  \; , \;
\xi_{\beta} = \left ( \begin{array}{c}
             \xi ^{1}_{\beta}  \\ \xi ^{2}_{\beta}
                                          \end{array} \right ) \; , \;
\eta_{\beta} = \left ( \begin{array}{c}
                 \eta _{\beta \dot{1}} \\ \eta _{\beta \dot{2}}
                                                        \end{array} \right )  \; , \;
\eqno(2.3a)
$$

\noindent
where ($\sigma ^{k}$ stand for Pauli two-by-two Pauli matrices, $k = 1,2,3$).
Introducing the  notation
$$
\sigma ^{\alpha } = \sigma ^{a} \; e^{\alpha }_{(a)} \; ,  \qquad
\bar{\sigma }^{\alpha }  = \bar{\sigma }^{a} \; e ^{\alpha } _{(a)}\; ,
$$
$$
B_{\alpha} = \left ( \begin{array}{ll}
\Sigma_{\alpha}  & 0 \\ 0  & \bar{\Sigma}_{\alpha}  \end{array} \right )
$$
$$
\Sigma _{\alpha } = {1 \over 2} \; \Sigma ^{ab} \; e^{\beta }_{(a)} \;
\nabla _{\alpha } (e_{(b)\beta }) \;  ,  \qquad
\bar{\Sigma } _{\alpha } = {1 \over 2} \; \bar{\Sigma }^{ab}
\; e^{\beta }_{(x)} \; \nabla _{\alpha } (e_{(b)\beta }) \;   ,
$$
$$
\Sigma ^{ab} = {1 \over 4}\; (\;  \bar{\sigma }^{a} \;
\sigma ^{b} \;  - \; \bar{\sigma }^{b} \; \sigma ^{a} \;)\; , \qquad
\bar{\Sigma }^{ab}={1  \over  4}  \; ( \; \sigma ^{a}\; \bar{\sigma }^{b}\; -
\;\sigma ^{b} \; \bar{\sigma }^{a}  \; ) \; ,
\eqno(2.3b)
$$

\noindent
instead of eqs. (2.1) one will obtains four equations\footnote{Those may seem somewhat more complex
that the above (2.1) and (2.2), but as will be seen later on it is not the case.}:
$$
\bar{\sigma}^{\alpha} (\nabla_{\alpha} + \bar{\Sigma}_{\alpha} - ig A_{\alpha} ) \eta +
$$
$$
+
\mu ( \sqrt{3} g^{\alpha \beta} + 2 \Sigma^{\alpha \beta} )
(\nabla_{\alpha} + \Sigma_{\alpha} - ig A_{\alpha}) \xi _{\beta} = M \xi \; ,
\eqno(2.4a)
$$
$$
\sigma^{\alpha} (\nabla_{\alpha} +  \Sigma_{\alpha} - ig A_{\alpha} ) \xi  +
$$
$$
+  \mu ( \sqrt{3} g^{\alpha \beta} + 2 \bar{\Sigma}^{\alpha \beta} )
(\nabla_{\alpha} + \bar{\Sigma}_{\alpha} - ig A_{\alpha}) \eta _{\beta} =  M \eta  \; ,
\eqno(2.4b)
$$
$$
\mu ( \sqrt{3} \delta_{\beta}^{\;\;\rho} - 2 \Sigma_{\beta}^{\;\;\rho})
(\nabla_{\rho} + \Sigma_{\rho} -igA_{\rho}) \xi = M \xi_{\beta} \; ,
\eqno(2.5a)
$$
$$
\mu ( \sqrt{3} \delta_{\beta}^{\;\;\rho} - 2 \bar{\Sigma}_{\beta}^{\;\;\rho})
(\nabla_{\rho} + \bar{\Sigma}_{\rho} -ig A_{\rho}) \eta = M \eta_{\beta} \; .
\eqno(2.5b)
$$

As long as the above equations (2.2) and (2.5) involve a tetrad manifestly, while the tetrad itself
at a given metric tensor  $g_{\alpha \beta}(x)$ is not uniquely defined:
$$
g^{kl}e_{(k)\alpha}
e_{(l)\beta}          = g_{\alpha \beta} \Longrightarrow \;\;
e'_{(a) \beta}   = L_{a}^{\;\; k}(x) e_{(k) \beta} \; ,
\eqno(2.6)
$$

\noindent  any two equations, associated respectively with
tetrads $e'_{(a) \beta}$ and $e_{(k) \beta}$, must be related to
each other by means of a local gauge transformation. We make the mechanism of
this gauge symmetry more explicit by seeing how these equations transform at a tetrad change.
To this end, one is to subject the constituents  the wave function to
{\em local } Lorentz transformations as prescribed by their spinor nature:
$$
\xi'  = B \; \xi  \; , \; \eta'  = [B^{+}]^{-1} \eta  \; , \;\;
$$
$$
\xi_{\beta}'  = B \; \xi_{\beta}  \; ,
\; \eta'_{\beta}  = (B^{+})^{-1} \eta _{\beta}  \; .
\eqno(2.7)
$$

\noindent Here a two-by-two matrix $B$ belongs to a special linear group $SL(2.C)$, universal covering of
the restricted Lorentz group $L^{\uparrow}_{+}$; symbol '+' stands for the hermitian conjugate.
To demonstrate the required symmetry manifestly, we are to study the following expressions\footnote{For a while,
for the sake of simplicity in the formulas we will omit the symbol of an external electromagnetic field $A_{\alpha}$.}:
$$
B \bar{\sigma}^{\alpha}  B^{+} \;  \; , \;\;
(B^{+})^{-1} \sigma^{\alpha } B^{-1}  \; , \;\;
\eqno(2.8a)
$$
$$
B \Sigma^{\alpha \beta} B^{-1}   \;  \; , \;\;
(B^{+})^{-1} \bar{\Sigma}^{\alpha \beta} B^{-1} \; \; , \;\;
\eqno(2.8b)
$$
$$
B (\nabla_{\alpha} + \Sigma_{\alpha} ) B^{-1}   \; \; , \;\;
(B^{+})^{-1} (\nabla_{\alpha} + \bar{\Sigma}_{\alpha} ) B^{+}  \;    \;  .
\eqno(2.8c)
$$

First consider (2.8a):
$$
B \bar{\sigma}^{\alpha}  B^{+} =  e_{(a)}^{\alpha} B \bar{\sigma}^{a} B^{+} \; , \;\;
(B^{+})^{-1} \sigma^{\alpha } B^{-1} = e_{(a)}^{\alpha} (B^{+})^{-1}  \sigma^{a} B^{-1} \; .
$$

\noindent From this, taking in mind the equalities\footnote{They are sometimes referred to as
the Dirac equation's  relativistic invariance conditions.} (their positive proof will be done below in Sec.3)
$$
B \bar{\sigma}^{a} B^{+} = \bar{\sigma}^{b} L_{b}^{\;\;a} \; , \;\;
(B^{+})^{-1}  \sigma^{a} B^{-1} = \sigma^{b}  L_{b}^{\;\;a} \; .
\eqno(2.9)
$$

\noindent where  $L_{b}^{\;\;a}$ designates four-by-four Lorentz matrix, related to the spinor matrix $B$
in (2.7), we come to
$$
B \bar{\sigma}^{\alpha}  B^{+} = \bar{\sigma}^{'\alpha} \; , \;\;
(B^{+})^{-1} \sigma^{\alpha } B^{-1} = \sigma ^{'\alpha}\;  .
\eqno(2.10)
$$

\noindent Here primed matrices are built on the base of the primed tetrad. Relationships (2.10) are just
 those we need.

For  (2.8b), with the use of eqs.  (2.10), one will  easily find
$$
B \Sigma^{\alpha \beta} B^{-1} = \Sigma^{'\alpha \beta}   \; , \;\;
(B^{+})^{-1} \bar{\Sigma}^{\alpha \beta} B^{-1} = \bar{\Sigma}^{'\alpha \beta} \; .
\eqno(2.11)
$$

Let us turn to  (2.8c). We are to prove two relationships:$$
B (\nabla_{\alpha} + \Sigma_{\alpha} ) B^{-1} = (\nabla_{\alpha} + \Sigma'_{\alpha} )   \; , \;\;
$$
$$
(B^{+})^{-1} (\nabla_{\alpha} + \bar{\Sigma}_{\alpha} ) B^{+} =
(\nabla_{\alpha} + \bar{\Sigma}'_{\alpha} )    \; ,
\eqno(2.12а)
$$

\noindent which are equivalent to the expected gauge transformation laws for 2-spinor connections:
$$
B \Sigma_{\alpha} B^{-1} + B \partial_{\alpha} B^{-1} = \Sigma'_{\alpha} \; , \;
$$
$$
(B^{+})^{-1}  \bar{\Sigma}_{\alpha} B^{+} +
(B^{+})^{-1}  \partial_{\alpha} B^{+} = \bar{\Sigma}'_{\alpha} \; .
\eqno(2.12b)
$$

\noindent  Here again primed connections  $\Sigma'_{\alpha}(x) $ and $\bar{\Sigma}'_{\alpha}(x) $
are constructed with the use of a primed tetrad. The validity of the laws  (2.12b)  will
be demonstrated in the next Sec. 3\footnote{First, it is technically  rather laborious task; second, it will
be useful time to introduce a number of facts and formulas on the theory of the Lorentz group and its
universal covering just in a manner convenient for exploiting in the present work.}.

Thus, the gauge invariance of the generally covariant Petras equation with respect to
tetrad local changes (2.7) belonging to the $SL(2.C)$ group, has been established. The set of primed
field functions, representing a fermion as observed in the $e_{(b)}^{'\beta}$-tetrad, obeys formally
the same equation (see (2.4) and (2.5) ) as the initial ones .
 In other words, the basic equation is form invariant under local gauge  (tetrad) transformations.
Of this symmetry fact  seems to be quite satisfactory, because the inspiration itself for any particle
field theory  has most often been found in underlying invariance principles. Thus, we in the
first place are interested in theories with definite behavior  under special relativity as well as under
general relativity.

In closing this Section else one addition may be done. In operating with
the covariant derivative  $D_{\alpha}$ we are very frequently to employ one auxiliary relation,
this is a useful time to prove it.  With this end in view, turn to the formula
$ S \gamma^{a} S^{-1} = \gamma^{b} L_{b}^{\;\;a} \;$\footnote{In the spinor representation
it is decomposed into eqs. (2.9).} and consider it for a infinitesimal Lorentz transformation;
then from the above it follows
$$
\sigma^{kl} \gamma^{a} - \gamma^{a} \sigma^{kl} =
\gamma^{b} (V^{kl})_{b}^{\;\;a} \; , \; (V^{kl})_{b}^{\;\;a} =
-g^{ka} \delta^{l}_{b} -  g^{la} \delta^{k}_{b}  \; .
$$

\noindent Next, allowing for the explicit form of the vector field generators $(V^{kl})_{b}^{\;\; a}$,
we arrive at the commutation relations
$$
\sigma^{kl} \gamma^{a} - \gamma^{a} \sigma^{kl} = \gamma^{k} g^{lk} - \gamma ^{l} g^{ka} \; .
\eqno(2.13)
$$

\noindent
Now, multiplying eq. (2.13) by a  tetrad-based expression
$
e_{(a)}^{\rho} \; {1 \over 2} e_{(k)}^{\sigma}    \nabla_{\beta}  e_{(l)\sigma } \; $ ,
we have got to
$$
\gamma^{\rho} \; B_{\beta}  -
B_{\beta} \; \gamma^{\rho}  =
\nabla_{\beta} \gamma^{\rho}   \; ,
\eqno(2.14)
$$

\noindent which is just that we need.
In turn, eq. (2.14) lead to the following commutation rule:
$$
D_{\beta} \gamma^{\rho} (x) = \gamma^{\rho} (x) D_{\beta}
\eqno(2.15)
$$

\noindent
In the spinor representation, eqs.  (2.14) and (2.15)  will be split into
$$
\bar{\sigma}^{\rho} \bar{\Sigma}_{\beta} - \Sigma_{\beta}\bar{\sigma}^{\rho} =
\nabla_{\beta} \bar{\sigma}^{\rho} \; , \;
$$
$$
\sigma^{\rho} \Sigma_{\beta} - \bar{\Sigma}_{\beta} \sigma^{\rho} =
\nabla_{\beta} \sigma^{\rho} \; , \;
\eqno(2.16)
$$
$$
(\nabla_{\beta} + \Sigma_{\beta})  \bar{\sigma}^{\rho} =
 \bar{\sigma}^{\rho} (\nabla_{\beta} + \bar{\Sigma}_{\beta}) \; , \;
$$
$$
(\nabla_{\beta} + \bar{\Sigma}_{\beta})  \sigma^{\rho} =
 \sigma^{\rho} (\nabla_{\beta} + \Sigma_{\beta}) \; ,
\eqno(2.17)
$$

\noindent take notice on  the placement of  the bar in these formulas.

\subsection*{3. $SL(2.C)$ group and  gauge properties of spinor
connections}

In this Section we will be interested in some technicalities of the Lorentz group theory,
and its universal covering  group $SL(2.C)$.
In the first place, we will need the following 'multiplication rules' for tow sets of Pauli matrices
$\sigma^{a}, \bar{\sigma}^{a}$:
$$
\bar{\sigma}^{a} \sigma^{b} \bar{\sigma}^{c} =
\bar{\sigma}^{a} g^{bc} - \bar{\sigma}^{b} g^{ac} + \bar{\sigma}^{c} g^{ab} - i
\epsilon^{abcd} \bar{\sigma}_{d} \; ,
$$
$$
\sigma^{a} \bar{\sigma}^{b} \sigma^{c} =
\sigma^{a} g^{bc} - \sigma^{b} g^{ac} + \sigma^{c} g^{ab} + i
\epsilon^{abcd} \sigma_{d} \; , \; \epsilon^{0123} = +1 \; .
\eqno(3.1)
$$

\noindent
They can be merged into one formula that is known in the literature as a multiplication rule for
four-by-four Dirac matrices [12]\footnote{However, the couple (3.1) should be understood as a more
fundamental then eq.  (3.2) .}:
$$
\gamma^{a} \gamma^{b} \gamma^{c} =
\gamma^{a} g^{bc} - \gamma^{b} g^{ac} + \gamma^{c} g^{ab} + i \gamma^{5}
\epsilon^{abcd} \gamma_{d} \; ,  \;
\gamma^{5} =
\left ( \begin{array}{cc}
-I &  0  \\ 0  &  + I
\end{array} \right ) \; .
\eqno(3.2)
$$

\noindent
Besides, we will need some formulas for the traces:
$$
sp ( \bar{\sigma}^{a} \sigma^{b}) = 2 g^{ab} \; , \;
sp ( \sigma^{a} \bar{\sigma}^{b}) = 2 g^{ab} \; , \;
$$
$$
sp (\bar{\sigma}^{a} \sigma^{b} \bar{\sigma}^{c} \sigma^{k} )=
2 ( g^{ab} g^{ck} - g^{ac} g^{bk} + g^{ak} g^{bc} -  i \epsilon^{abck} )\; , \;
$$
$$
sp (\sigma^{a} \bar{\sigma}^{b} \sigma^{c} \bar{\sigma}^{k} )=
2 ( g^{ab} g^{ck} - g^{ac} g^{bk} + g^{ak} g^{bc} +  i \epsilon^{abck}) \; .
\eqno(3.3)
$$

Now, let us turn again  to eqs. (2.9):
$$
B \bar{\sigma}^{a} B^{+} = \bar{\sigma}^{b} L_{b}^{\;\;a} \; , \;\;
(B^{+})^{-1}  \sigma^{a} B^{-1} = \sigma^{b}  L_{b}^{\;\;a} \; .
\eqno(3.4)
$$

\noindent  Now, we will show that these are indeed valid, at the same time will establish an
explicit representation for a Lorentz four-by-four matrix  as a function of parameters
$k_{a}$ and $k_{a}^{*}$.

With the use of trace formulas  (3.2),  eqs. (3.4)  can be easily solved under the matrix $L_{b}^{\;\;a}$:
$$
L_{b}^{\;\;a} ={1 \over 2} \; sp \; [ \sigma_{b}  B  \bar{\sigma}^{a} B^{+}) \; ,
$$
$$
L_{b}^{\;\;a} ={1 \over 2} \; sp \; [ \bar{\sigma}_{b}  (B^{+})^{-1} \sigma^{a} B^{-1}) \; .
\eqno(3.5)
$$

\noindent In the following it will be convenient to utilize a natural parameterization of spinor
two-by-two matrices through 4-dimensional complex vector\footnote{Of course,
 we could have chosen other
(more physical one: for example, considering a pure rotation followed by a pure boost,
 but it will be useful to have an explicit form at hand.}
$$
B(k) = \; \sigma ^{a} \; k_{a}  \;  , \;
\det  B =  k^{2}_{0} \; - \; k^{2}_{j}  = + 1 \;  ,
$$
$$
B^{+}(k) = B(k^{*}) \; , \qquad B^{-1}(k) = B(\bar{k})\; , \qquad
\bar{k} = (k _{0},\; - k_{j}) \; .
\eqno(3.6)
$$

\noindent Taking this notation in mind, from eqs.  (3.5) it follows
$$
L_{b}^{\;\;a} =
{1 \over 2} \; sp \; [\; \sigma_{b}  B(k)  \bar{\sigma}^{a} B(k^{*}) ] =
{1 \over 2} \; sp \; [\sigma_{b}  \sigma^{n} k_{n}
\bar{\sigma}^{a} \sigma^{m} k^{*}_{m} \; ] \; ,
\eqno(3.7a)
$$
$$
L_{b}^{\;\;a} = {1 \over 2} \; sp \; [ \; \bar{\sigma}_{b}  B(\bar{k}^{*})
\sigma^{a} B(\bar{k}) ] =
{1 \over 2} \; sp \; [ \bar{\sigma}_{b} \bar{\sigma}^{m} k^{*}_{m}
\sigma^{a} \bar{\sigma}^{n} k_{n} \; ]  \; .
\eqno(3.7b)
$$

\noindent Introducing special Kronecker's symbol
$$
\bar{\delta}^{c}_{b} = \left \{ \begin{array}{l}
0 , c \neq b  \; ,\\
+1 , c =b = 0 \; , \\
-1, c= b = 1= 2=3 \; ,  \end{array} \right.
$$

\noindent  eqs. (3.7) read as
$$
L_{b}^{\;\;a} = \bar{\delta}^{c}_{b} \;
{1 \over 2} \; sp \; [ \; \bar{\sigma}_{c}  \sigma^{n}
\bar{\sigma}^{a} \sigma^{m} \; ] \; k_{n}  k^{*}_{m}   \; ,
\eqno(3.8a)
$$
$$
L_{b}^{\;\;a} = \bar{\delta}_{b}^{c} \;
{1 \over 2} \; sp \; [ \; \sigma_{c} \bar{\sigma}^{m}
\sigma^{a} \bar{\sigma}^{n} \; ] \;  k_{n} k^{*}_{m}    \; .
\eqno(3.8b)
$$

\noindent Next, with the use of the trace formulas (3.3), one can easily find
$$
L_{b}^{\;\;a} = \bar{\delta}^{c}_{b} \;
{1 \over 2} \; sp \; [ \; \bar{\sigma}_{c}  \sigma^{n}
\bar{\sigma}^{a} \sigma^{m} \; ] \; k_{n}  k^{*}_{m}
=
\bar{\delta}^{c}_{b} \;
{1 \over 2} \; sp \; [ \; \sigma_{c} \bar{\sigma}^{m}
\sigma^{a} \bar{\sigma}^{n} \; ] \;  k_{n} k^{*}_{m} =
$$
$$
= \bar{\delta}^{c}_{b} \;
[ \;  - \delta^{a}_{c} k^{n} k_{n}^{*}  +  k_{c} k^{a*} + k^{*}_{c} k^{a} +
i \epsilon_{c}^{\;\; anm} k_{n} k^{*}_{m} \; ] \; .
\eqno(3.9)
$$

\noindent
By this relation we are supplied with a parametrization of
Lorentz four-by-four matrix $L_{b}^{\;\;c}$ in terms of two
complex vectors  $ k_{n}$ and $ k^{*}_{n}$.

It can be straightforwardly shown that the matrix defined by (3.9) possesses the orthogonality property,
what is expected for any Lorentz transformation:
 $$
 L^{\;\; b}_{a}(k,\; k^{*}) = g^{bc} \; L^{\;\; d}_{c} (\bar{k},\; \bar{k}^{*})  \; g_{da} \; .
\eqno(3.10)
$$

\noindent Indeed, eq. (3.10) can be rewritten as
$$
\bar{\delta }^{c}_{a} \; [ \;- \delta ^{b}_{c} k^{n} k^{*}_{n}
+ k _{c} k^{b*} + k^{*}_{c} k^{b} + i \epsilon ^{\;\;bnm}_{c}  k_{n} k^{*}_{m} \; ]  =
$$
$$
=
\bar{g}^{bc} \; [ \; - \delta ^{d}_{c} \bar{k} ^{n}
 \bar{k}^{*}_{n}  +  \bar{k}_{c} \bar{k}^{d*} +
\bar{k}^{*}_{c} \bar{k} ^{d} + i
\epsilon  ^{\;\;dnm}_{c}  \bar{k}_{n} \bar{k}^{*}_{m}
\; ] \;  g _{da}
$$

\noindent  where the notation   $\bar{g}^{bc} = g^{ba} \bar{\delta }^{c}_{a}$ is used.
From the above it follows
$$
( \; \bar{\delta }^{c}_{a} \; \epsilon ^{\;\;bnm}_{c}
\; ) \; k_{n} \; k^{*}_{m}  =  ( \; \bar{g}^{bc} \;
 \epsilon ^{\;\;dnm}_{c} \; g_{da} \; ) \;
\bar{k}_{n} \; \bar{k}^{*}_{m} \; .
$$

\noindent This identity is valid by a direct verifying that it holds for every choice of indices.

Next, we are to check that it is orthochronic Lorentz transformation, that is
$L^{\;\; 0}_{0}(k,\; k^{*}) \ge  + 1$ . Indeed, an explicit form for  $L^{\;\; 0}_{0}$ is
$
L^{\;\;0}_{0} = (\; k_{0} \; k^{*}_{o} \;  + \; k_{j}\;  k^{*}_{j}\; ) \; ;
$
from this, with the use of inequality
$$
(\mid Z_{0}\mid+ \mid Z_{1}  \mid  + \mid Z_{2}\mid  +
 \mid Z_{3} \mid ) \ge \mid Z_{0} + Z_{1} +
Z_{2} + Z_{3} \mid    \; ,
$$

\noindent
at  $Z_{0}= k_{0} \; k_{0}, \; Z_{1}= - k_{1}\; k_{1} , \;
Z_{2} = - k_{2} \; k_{2} , \;  Z_{3} = - k_{3} \; k_{3}$, and taking into account
$det \; B(k) = 1 $, we obtain
$$
L^{\;\; 0}_{0} \; \ge \; \mid k_{0}\; k_{0}\; - k_{j} \; k_{j} \mid \; =
\;+\;1 \; .
$$

\noindent
Else one property of the  Lorentz matrix according to (3.9), $\det  L(k,\;k^{*}) = +1$,  should be proved
too;  this is simple (though laborious) task and  its solution is omitted.

Now we are to verify  the formulas  (3.12b). Obviously, it
suffices to consider in detail only one of them; for definiteness let it be the first one
$$
B \Sigma_{\alpha} B^{-1} + B \partial_{\alpha} B^{-1} = \Sigma'_{\alpha} \; .
\eqno(3.11a)
$$

\noindent For the term
$$
B \; \Sigma_{\alpha} \; B^{-1} =
{1 \over 2} \;B
\Sigma^{ab}   B^{-1} \;  e_{(a)}^{\;\beta} (\nabla_{\alpha} e_{(b)\beta})   \;
,
$$

\noindent expressing the tetrad  $e_{(b)\beta}$  in terms of primed $e'_{(b)\beta}$, we have
$$
B \; \Sigma_{\alpha} \; B^{-1} = {1 \over 2} \;( B \; \Sigma^{ab} B^{-1} )
\; (L^{-1})_{a}^{\;\;k} e_{(k)}^{'\;\beta} \;
[ \; \nabla_{\alpha} (L^{-1})_{b}^{\;\;l} e_{(l)\beta} \; ] \; ,
$$

\noindent and further
$$
B \; \Sigma_{\alpha} \; B^{-1} =
{1 \over 2} \;( B \; \Sigma^{ab} B^{-1} )
\;
\; \left [ \;  (L^{-1})_{a}^{\;\;k} e_{(k)}^{'\;\beta }\;
\; [ \; ( \partial _{\alpha} (L^{-1})_{b}^{\;\;l} \; )  \;
e'_{(l)\beta}  +  (L^{-1})_{b}^{\;\;l} \nabla_{\alpha} e'_{(l)\beta} \; ] \right ] =
$$
$$
=
{1 \over 2} \;( B \; \Sigma^{ab} B^{-1} ) \;  \left [
 \; (L^{-1})_{a}^{\;\;k} \;
 (\partial _{\alpha} (L^{-1})_{b}^{\;\;l})\;
e_{(k)}^{'\;\beta} \;  e'_{(l)\beta}   +
(L^{-1})_{a}^{\;\;k}  (L^{-1})_{b}^{\;\;l} \;
e_{(k)}^{'\;\beta} \;  \nabla_{\alpha} e'_{(l)\beta} \right ] \; .
$$

\noindent Thus, we have arrived to
$$
B \; \Sigma_{\alpha} \; B^{-1} =
{1 \over 2} \;( B \; \Sigma^{ab} B^{-1} ) \; \times
$$
$$
\times \;
\left [
(L^{-1})_{a}^{\;\;k}
(\partial_{\alpha} (L^{-1})_{b}^{\;\;l})\; g_{kl} +
(L^{-1})_{a}^{\;\;k}  (L^{-1})_{b}^{\;\;l} \;
e_{(k)}^{'\;\beta} \;  \nabla_{\alpha} e'_{(l)\beta} \right ]  \; .
\eqno(3.11b)
$$

\noindent Next, taking into account the identity, simple consequence
of (3.9) ),
$$
B \Sigma^{ab} B^{-1} = \Sigma^{mn} L_{m}^{\;\;\;a} L_{n}^{\;\;b} \; ,
$$

\noindent  from (3.11b) it follows
$$
B \; \Sigma_{\alpha} \; B^{-1} =
 {1 \over 2} \; \Sigma^{mn} L_{n}^{\;\;b}
\partial _{\alpha} (L^{-1})_{bm}  +  \Sigma'_{\alpha}  \; .
\eqno(3.11c)
$$

\noindent And finally, allowing for  eq. (3.11c), from  (3.11a) we get to
$$
B \partial_{\alpha} B^{-1}
- {1 \over 2} \; \Sigma^{mn}
 L_{m}^{\;\;b} \partial _{\alpha} (L^{-1})_{bn} = 0 \; .
\eqno(3.12a)
$$

\noindent
Much in the same manner, from second relation in (3.9b) it can be found
$$
(B^{+})^{-1} \partial_{\alpha} B^{+}
- {1 \over 2} \; \bar{\Sigma}^{mn}  L_{m}^{\;\;b} \partial _{\alpha} (L^{-1})_{bn} = 0  \; .
\eqno(3.12b)
$$

\noindent
It is convenient to alter the equalities  (3.12a,b)  into
$$
B \partial_{\alpha} B^{-1} - {1 \over 4} \;
(\bar{\sigma}^{m} L_{m}^{\;\;b} )   \partial _{\alpha}
(\sigma^{n}   L^{-1})_{bn}  ) = 0  \; ,
\eqno(3.13a)
$$
$$
(B^{+})^{-1}  \partial_{\alpha} B^{+} - {1 \over 4} \;
( \sigma^{m} L_{m}^{\;\;b} )   \partial _{\alpha}
(\bar{\sigma}^{n}   L^{-1})_{bn}  ) = 0  \; .
\eqno(3.13b)
$$

\noindent From where, with  the use of eq. (3.9),  we get to
$$
B \partial_{\alpha} B^{-1} - {1 \over 4} \;
B \bar{\sigma}^{b} B^{+}   \;  \partial _{\alpha}
\; [\;  (B^{+})^{-1}\sigma_{b} B^{-1} \; ] = 0  \; ,
$$
$$
(B^{+})^{-1} \partial_{\alpha} B^{+} - {1 \over 4} \;
 (B^{+})^{-1} \bar{\sigma}^{b} B^{-1}   \;  \partial _{\alpha}
\; [\;  (B \sigma_{b} B^{+}\;  ] = 0  \; ,
$$

\noindent which are equivalent to
$$
-{1 \over 4} \;
B \; \{ \; \bar{\sigma}^{a} \; [ \;
 B^{+} \partial_{\alpha} (B^{+})^{-1} \; ] \;  \sigma_{b}
\;  \} \; B^{-1} = 0 \; ,
\eqno(3.14a)
$$
$$
-{1 \over 4} \;
(B^{+})^{-1} \;  \{ \; {\sigma}^{b} \; [ \;
B^{-1} \partial_{\alpha} B \; ]  \;
\bar{\sigma}_{b} \; \} \;  B^{+} = 0 \; .
\eqno(3.14b)
$$

\noindent
Eqs. (3.14a,b) involve  two identities
ноль:
$$
\bar{\sigma}^{a} \; [\;  B^{+} \partial_{\alpha} (B^{+})^{-1} \; ] \;\sigma_{b}
 \equiv  0 \; , \;\;
{\sigma}^{b} \; [ \; B^{-1} \partial_{\alpha} B \; ] \;
\bar{\sigma}_{b} \equiv 0 \; ,
\eqno(3.15)
$$

\noindent which, in their turn,  follow from the known formulas
(indices $p,s$ take on the values
$1,2,3$)
$$
B^{+} \partial_{\alpha} (B^{+})^{-1}  =
\sigma_{j}\;  ( \;  k_{0} \; \partial_{\alpha} k_{j} - k_{j} \;  \partial_{\alpha} \;
k_{0} +  i \epsilon_{jps} k_{p} \; \partial_{\alpha} k_{c} ) \; ,
$$
$$
B^{-1} \partial_{\alpha} B  = -
\sigma_{j} \; ( \;  k_{0} \; \partial_{\alpha} k_{j} - k_{j} \;
 \partial_{\alpha} \;  k_{0} +
i \epsilon_{jps} k_{p}\;  \partial_{\alpha} k_{c} ) \; ,
\eqno(3.16)
$$

\noindent  and the evident identities
$$
\bar{\sigma}^{b} \sigma_{j} \sigma_{b}  = 0 \; , \;
\sigma^{b} \sigma_{j} \bar{\sigma}_{b} = 0 \; .
$$

\noindent
So, the gauge transformation laws for Infeld-van der Vaerden connections
$$
B \Sigma_{\alpha} B^{-1} + B \partial_{\alpha} B^{-1} = \Sigma'_{\alpha} \; , \;
$$
$$
(B^{+})^{-1}  \bar{\Sigma}_{\alpha} B^{+} +
(B^{+})^{-1}  \partial_{\alpha} B^{+} = \bar{\Sigma}'_{\alpha} \;
\eqno(3.17)
$$

\noindent are  proven\footnote{It should be   mentioned that these gauge properties  are commonly
known, all justification for else one treatment of those is that their  proof is given with the use of finite
transformations.}.

\subsection*{4. Invariant form matrix and conserved current}

Section 4 will provide a framework for addressing  the fundamental
dynamical questions: what will the Lagrange function of  the  generalized fermion
look like and in what manner  the various bilinear combinations, physically observable quantities
are to be constructed.

With this end in view, we are to consider the question about a matrix of invariant bilinear
 form\footnote{Sometimes it is referred to as the hermitian matrix.} and a generalized  conserved current.
It is convenient to start with the  original equations in the form (see (3.1) and (3.2))
$$
\gamma^{\alpha} D_{\alpha} \Psi  + \mu  \;
( \sqrt{3} g^{\alpha \beta}  + 2 \sigma^{\alpha \beta}  )  \;
          D_{\alpha} \Psi _{\beta}  - M \Psi  = 0 \; ,
\eqno(4.1)
$$
$$
\mu \; (\sqrt{3} \delta_{\beta}^{\;\;\alpha} -
 2 \sigma_{\beta}^{\;\;\alpha} ) \; D_{\alpha} \Psi  - M \Psi_{\beta}  = 0 \;
 ,
\eqno(4.2)
$$

 \noindent here  $ D_{\alpha} = \nabla_{\alpha} + B_{\alpha}  - ig A_{\alpha} $.

First, let us find an equation conjugate to eq. (4.1). Acting on eq.  (4.1) by the  hermitian conjugate
operation, we get to
$$
\Psi^{+}  \; \hat{D}_{\alpha}^{+}  \; \gamma^{\alpha +}  +  \mu\;  \Psi_{\beta}^{+} \;
\hat{D}_{\alpha}^{+}  \;
(\; \sqrt{3} \;  g^{\alpha\beta} + 2 \sigma^{\alpha \beta +} \; )  + M \; \Psi ^{+} = 0 \; .
\eqno(4.3a)
$$

\noindent Here $\hat{D}_{\alpha}$ stands for an derivative operation on the left
$$
\hat{D}_{\alpha}^{+} = \stackrel{\leftarrow}{\nabla}_{\alpha} +
B^{+}_{\alpha}   + ig A_{\alpha}  \; .
\eqno(4.3b)
$$

\noindent
For definiteness it is convenient to assume the use of the Weyl's spinor
representation of the Dirac matrices\footnote{In Supplement ... , the employing of  an arbitrary
Dirac's matrix basis in the context of the given problem will be described in some
detail.},in which the two following properties  are
readily  verified
$$
\gamma^{0} \;  \gamma^{\alpha + }(x) \; \gamma^{0} = +  \gamma^{\alpha}(x)  \; , \;\;
\gamma^{0} \; \sigma^{\alpha \beta + } (x) \;  \gamma^{0} = -
\sigma^{\alpha \beta }(x)  \; , \;\;
\eqno(4.4)
$$

\noindent With (4.4), from  (4.3a) it follows
$$
\Psi^{+} \gamma^{0} \stackrel{\leftarrow}{D}_{\alpha} \gamma^{\alpha} +
\mu \Psi_{\beta}^{+} \gamma^{0} \stackrel{\leftarrow}{D}_{\alpha}
( \sqrt{3} g^{\alpha\beta} - 2 \sigma^{\alpha \beta} )  + M \Psi ^{+} \gamma^{0} = 0 \; ;
\eqno(4.5a)
$$

\noindent where  $\stackrel{\leftarrow}{D}_{\alpha} $ stands for a
new derivative operation on the left:
$$
\stackrel{\leftarrow}{D}_{\alpha}
= \stackrel{\leftarrow}{\nabla}_{\alpha} -
B_{\alpha}   + ig A_{\alpha}  \; .
\eqno(4.5b)
$$

\noindent It may be noted that the $\stackrel{\leftarrow}{\nabla}_{\alpha}$
obeys the commutation relation similar to (3.14b)
$$
\stackrel{\leftarrow}{D}_{\beta} \;  \gamma^{\rho} (x) =
\gamma^{\rho}(x) \;  \stackrel{\leftarrow}{D}_{\beta} \; .
$$

As usual, for a combination  $\Psi^{+} \gamma^{0}$ the notation $\bar{\Psi}$ will be used, and
the $\bar{\Psi}$-function will be interpreted as conjugate to the $\Psi$. However, a similar
combination as applied to the vector-bispinor constituent:  $\Psi_{\beta}^{+} \gamma^{0}$,
cannon be understood in a similar  way as a conjugate to the $\Psi_{\beta}$.
But the right answer to this problem follows from requeremant that
the relation
$$
(\; \Psi_{\beta}^{+} \gamma^{0}  \; \eta ^{\beta}_{\;\;\nu} \; ) \;
(\eta^{-1})^{\nu}_{\;\;\rho} \stackrel{\leftarrow}{D}_{\alpha}
\; ( \sqrt{3} g^{\alpha \rho} - 2 \sigma^{\alpha \rho} ) =
\bar{\Psi}_{\nu} \; \stackrel{\leftarrow}{D}_{\alpha}  \;
( \; \sqrt{3} g^{\alpha \nu} + 2 \sigma^{\alpha \nu} \;  ) \; .
$$

\noindent be hold. The $\eta$-matrix might be identified by the formula
$$
(\eta^{-1})^{\nu}_{\;\;\rho}(x) = F  \; \delta^{\nu}_{\;\; \rho} + G \; \sigma^{\nu}_{\;\;\rho}(x) \; ,
$$

\noindent where  $F$ and $G$ are yet not-fixed numbers.  From  the relation
$$
(\; F \delta^{\nu}_{\;\; \rho} + G \sigma^{\nu}_{\;\;\rho} \; )
\; (\; \sqrt{3} g^{\rho \alpha }  + 2 \sigma^{\rho \alpha } \;  )  =
( \;  \sqrt{3} g^{\nu \alpha } - 2 \sigma^{\nu \alpha } \; )
$$

\noindent we obtain equations for  $F$ and $G$
$$
{ \sqrt{3} \over 2} F - {2 F + \sqrt{3} G \over 4 } + { G \over 4 } = {\sqrt{3} \over 2}
+ {1 \over 2} \; ,
$$
$$
{ \sqrt{3} \over 2} F + {2 F + \sqrt{3} G \over 4 } + { G \over 4 } = {\sqrt{3} \over 2}
- {1 \over 2} \; .
$$

\noindent Their explicit solution is   $F = 1 + \sqrt{3} \; , \; G = - 2 $.
Therefore, the  $\eta^{-1}$-matrix is given by
$$
(\eta^{-1})^{\nu}_{\;\;\rho}(x) =
(1 + \sqrt{3}) \; \delta^{\nu}_{\;\; \rho} - 2 \; \sigma^{\nu}_{\;\;\rho}(x) \; .
\eqno(4.6a)
$$
\noindent Remember that this matrix is identified by
$$
(\eta^{-1})^{\nu}_{\;\;\rho} \;
( \;  \sqrt{3} \; g^{\rho \alpha}  + 2 \; \sigma^{\rho  \alpha } \; )  =
( \;  \sqrt{3} \; g^{\nu \alpha } - 2 \; \sigma^{\nu \alpha } \; ) \; .
\eqno(4.6b)
$$

\noindent It is readily verified the expression for $\eta$
$$
\eta^{\beta}_{\;\;\nu}(x) =
(1 - \sqrt{3}) \; \delta^{\beta}_{\;\;\nu} - 2 \; \sigma^{\beta }_{\;\;\nu }(x) \; .
\eqno(4.7)
$$

\noindent So,  a set of functions conjugate to the  $\Psi (x), \Psi_{\beta}(x)$
is defined as below
$$
\bar{\Psi} = \Psi ^{+} \; \gamma^{0} \; , \qquad
\bar{\Psi}_{\nu}  = \Psi ^{+}_{\beta} \; \gamma^{0}  \;
\eta^{\beta}_{\;\;\nu}(x) \; .
\eqno(4.8)
$$

\noindent  An equation conjugate to eq. (4.1) has the form  (see  (4.1))
$$
\bar{\Psi} \stackrel{\leftarrow}{D}_{\alpha}
\; \gamma^{\alpha} +
\mu \; \bar{\Psi}_{\beta}
\stackrel{\leftarrow}{D}_{\alpha} \;
(\; \sqrt{3} g^{\alpha\beta} + 2 \sigma^{\alpha \beta} \; )  + M \; \bar{\Psi} = 0 \; .
\eqno(4.9)
$$

Now, we are to obtain a second  conjugate  equation (see (4.2). To this end, acting on eq. (4.2) the
hermitian conjugate operation, we get to
$$
\mu \; \Psi^{+} \gamma^{0} \stackrel{\leftarrow}{D}_{\rho}  \;
( \; \sqrt{3} \delta_{\beta}^{\;\;\rho} +
2 \sigma_{\beta}^{\;\;\rho}\; )  +  M \; \Psi^{+}_{\beta} \; \gamma^{0}  = 0 \; .
$$

\noindent Next, multiplying that by $\eta^{\beta}_{\;\;\alpha}(x)$ from the left,
and taking into account the relation (this one can be derived by a direct calculation)
$$
(\; \sqrt{3} \; \delta^{\rho}_{\beta} - 2 \; \sigma_{\;\;\beta}^{\rho} )\;
\eta^{\beta}_{\;\;\alpha } =
\sqrt{3} \; \delta^{\rho}_{\alpha}  + 2 \; \sigma^{\rho}_{\;\;\alpha} \; ,
\eqno(4.10a)
$$

\noindent we get the result  we need
$$
\mu\; \bar{\Psi} \stackrel{\leftarrow}{D}_{\rho}
(\sqrt{3} \; \delta_{\alpha}^{\rho} -
2\;  \sigma_{\alpha}^{\;\;\rho} )  +  M \; \bar{\Psi}_{\alpha}   = 0 \; .
\eqno(4.10b)
$$

So, the full system of conjugate equations is as follows:
$$
\bar{\Psi} \stackrel{\leftarrow}{D}_{\alpha}
\gamma^{\alpha} +
\mu \; \bar{\Psi}_{\beta}
\stackrel{\leftarrow}{D}_{\alpha}
\; (\;  \sqrt{3} g^{\alpha\beta} + 2 \sigma^{\alpha \beta} \; )  + M \; \bar{\Psi} = 0 \; ,
\eqno(4.11a)
$$
$$
\mu \; \bar{\Psi} \stackrel{\leftarrow}{D}_{\alpha} \;
( \; \sqrt{3} \delta_{\beta} ^{\alpha} -
2 \sigma_{\beta}^{\;\;\alpha} \; )  +  M \; \bar{\Psi}_{\beta}   = 0 \; .
\eqno(4.11b)
$$

Now, we are ready to derive a conserved current  law. To this end,
multiplying  both eq.  (4.1) by $\bar{\Psi}$ and eq. (4.2)  by
$\bar{\Psi}^{\beta}$ from the left, and similarly multiplying eq. (4.11a)by $\Psi$ from the right as well as
eq. (4.11b) by $\Psi^{\beta}$ from the right, and summing the results, we  get to
$$
[\;  \bar{\Psi} \gamma^{\alpha} \stackrel{\rightarrow}{D} _{\alpha} \; \Psi  +
\bar{\Psi} \stackrel{\leftarrow}{D} _{\alpha} \;  \gamma^{\alpha}  \Psi \; ]  \; +
$$
$$
+ \mu \;[\;  \bar{\Psi} \;
(\; \sqrt{3} g^{\alpha \beta} + 2 \sigma^{\alpha \beta} \; ) \;
\stackrel{\rightarrow}{D} _{\alpha}   \Psi_{\beta}  +
\bar{\Psi} \stackrel{\leftarrow}{D} _{\alpha} \; (\; \sqrt{3} g^{\alpha \beta}
+ 2  \sigma^{\alpha \beta} \; ) \; \Psi_{\beta} \; ]  \; +
$$
$$
+ \mu \; [ \;
\bar{\Psi}_{\beta} \;  ( \; \sqrt{3} g ^{\beta \alpha } - 2
 \sigma^{ \beta \alpha}\; )  \;
\stackrel{\rightarrow}{D} _{\alpha}  \Psi \;  +\;
\bar{\Psi}_{\beta}  \stackrel{\leftarrow}{D} _{\alpha}   \;
 ( \; \sqrt{3} g^{\beta \alpha } - 2 \sigma^{\beta \alpha }\;  )\; ]  = 0 \; .
\eqno(4.12)
$$

The first term in  square brackets in (4.12), with the use of the above commutation relation (3.14a),
will reads as
$$
 \bar{\Psi} \gamma^{\alpha} \stackrel{\rightarrow}{D} _{\alpha} \Psi  +
\bar{\Psi} \stackrel{\leftarrow}{D} _{\alpha}  \gamma^{\alpha}  \Psi   =
$$
$$
\bar{\Psi} \stackrel{\leftarrow}{\nabla}_{\alpha}  \gamma^{\alpha} +
\bar{\Psi}\;  ( \gamma^{\alpha} B_{\alpha} - B_{\alpha} \gamma^{\alpha}) \; \Psi +
\bar{\Psi} \gamma^{\alpha} \stackrel{\rightarrow}{\nabla}_{\alpha} \Psi =
\nabla_{\alpha}  ( \bar{\Psi} \gamma^{\alpha} \Psi ) \; .
\eqno(4.13)
$$

This formula exhibits a notable feature: it provides a rule for acting the conventional covariant
derivative $\nabla_{\alpha}$ on a generally covariant vector constructed as a bilinear combination
from the field  $\Psi$ and a conjugate field  $\bar{\Psi}$. According to (4.13), one may replace the
the action of the $\nabla_{\alpha}$ by the action by  $\stackrel{\leftarrow}{D}_{\alpha} $ from the right
and by the action by $\stackrel{\rightarrow}{D}_{\alpha} $ from the left
on $\bar{\Psi}$ and  $\Psi$ respectively, and ignoring completely
the general covariant nature of the matrix  $\gamma^{\alpha}(x)$ in the middle.
This fact is representative one, being an example of a general rule.

Actually, let us consider, for instance, the following expression
$
\nabla^{\alpha} [ \; \bar{\Psi} \gamma^{\rho} \gamma^{\sigma} \Xi \; ]
\; $\footnote{Here $\Xi$ is an arbitrary function with transformation properties that
coincides with those for the  $\Psi $.}.
Rewriting it in the form
$$
\nabla^{\alpha} [ \; \bar{\Psi} \gamma^{\rho} \gamma^{\sigma} \Xi \; ] =
[ ( \partial_{\alpha} \bar{\Psi} )  \gamma^{\rho} \gamma^{\sigma} \Xi  +
\bar{\Psi} (\nabla_{\alpha} \gamma^{\rho} )  \gamma^{\sigma} \Xi +
$$
$$
+  \bar{\Phi} \gamma^{\rho}( \nabla_{\alpha} \gamma^{\sigma} )  \Xi  +
\bar{\Psi} \gamma^{\rho}   \gamma^{\sigma}
\; (\partial_{\alpha} \Xi ) ]\; ,
$$

\noindent
and replacing the covariant derivatives over  $\gamma$-matrices as prescribed by (3.14a), we get
$$
\nabla^{\alpha}\; [ \; \bar{\Psi} \gamma^{\rho} \gamma^{\sigma} \; \Xi \; ] =
$$
$$
= ( \partial_{\alpha} \bar{\Psi}) \;  \gamma^{\rho} \gamma^{\sigma} \; \Xi  +
\bar{\Psi} \;  (\gamma^{\rho} B_{\alpha}  - B_{\alpha} \gamma^{\rho}  ) \;
\gamma^{\sigma} \; \Xi \;  +
$$
$$
+ \bar{\Psi} \gamma^{\rho} \; (\gamma^{\sigma} B_{\alpha} - B_{\alpha} \gamma^{\sigma}  )\;  \Xi  +
\bar{\Psi} \gamma^{\rho} \gamma^{\sigma} \;  ( \partial_{\alpha} \Xi ) \; .
$$

\noindent Here two terms involving the product
$\gamma^{\rho} B_{\alpha}\gamma^{\sigma}$ cancel each other, and remaining expression can be
read as
$$
\nabla^{\alpha} \; ( \;  \bar{\Psi} \gamma^{\rho} \gamma^{\sigma} \Xi \; ) \;
 =  \bar{\Psi} \stackrel{\leftarrow}{D}_{\alpha} \;  \gamma^{\rho}
 \gamma^{\sigma} \Xi +
\bar{\Psi}  \gamma^{\rho} \gamma^{\sigma}     \;
\stackrel{\rightarrow}{D}_{\alpha} \Xi  \; .
\eqno(4.14a)
$$

\noindent Relation  (4.14a) is equivalent to
$$
\nabla_{\alpha} (  \gamma^{\rho} \gamma^{\sigma} ) =
- B_{\alpha}  \;  \gamma^{\rho}   \gamma^{\sigma}  +
\gamma^{\rho} \gamma^{\sigma}  \; B_{\alpha} \; .
\eqno(4.14b)
$$

Extension to a case of bilinear combination  of arbitrary tensor structure  is straightforward by induction.
Indeed, let $\gamma^{(n)}$ be
$$
\gamma^{(n)} = \gamma^{\rho_{1}}  \gamma^{\rho_{2}}
... \gamma^{\rho_{n}}  \; , \;\; и \;\;
\nabla_{\alpha}  \gamma^{(n)}  = \gamma^{(n)} B_{\alpha} -
B_{\alpha} \gamma^{(n)} \; ;
\eqno(4.15a)
$$

\noindent  then $\gamma^{(n+1)} = \gamma^{(n)} \gamma^{\rho} $ obeys
$$
\nabla_{\alpha} \gamma^{(n+1)} = - B_{\alpha} \gamma^{(n+1)} + \gamma^{(n+1)} B_{\alpha} \; .
\eqno(4.15б)
$$

\noindent
From eqs. (4.15) one can get the commutation relations we need
$$
\gamma^{(n) }  \stackrel{\rightarrow}{D_{\sigma}} =
         \stackrel{\rightarrow}{D_{\sigma}} \gamma^{(n)} \; , \;\;
\gamma^{(n)} \stackrel{\leftarrow}{D_{\sigma}} =
         \stackrel{\leftarrow}{D_{\sigma}} \gamma^{(n)}\; .
\eqno(4.16)
$$

\noindent or in an equivalent form
$$
\nabla_{\alpha} [ \bar{\Psi} \gamma^{(n)}(x) \Psi ] =
\bar{\Psi}   \stackrel{\leftarrow}{D_{\sigma}} \gamma^{(n)}(x) \Psi +
\bar{\Psi}   \gamma^{(n)}(x)  \stackrel{\rightarrow}{D_{\sigma}} \Psi \; .
\eqno(4.17)
$$

With the use of the established rules  (4.17), eq. (4.12) reads as a  conserved current law
$$
\nabla_{\alpha} J^{\alpha}(x) = 0 \; , \;
$$
$$
J^{\alpha}  =
\bar{\Psi} \gamma^{\alpha}  \Psi  +
\mu \;[ \bar{\Psi} ( \sqrt{3} g^{\alpha \beta} + 2 \sigma^{\alpha \beta} )
\Psi_{\beta}   +
\bar{\Psi}_{\beta}  (\sqrt{3} g ^{\beta \alpha } - 2 \sigma^{\beta \alpha})
 \Psi ]  \; .
\eqno(4.18)
$$

In connection with eq. (4.18),  a couple of fact should one discussed.
Really, note that the  first term in (4.18) is real-valued. So must be the remaining part of the given  current.
We are do demonstrate that second and third terms in (4.180 are complex conjugate to each other.
To this end, let as look at an expression conjugate to the second term (remember eqs. (4.4)):
$$
[ \; \bar{\Psi} \; ( \sqrt{3} g^{\alpha \beta} + 2 \sigma^{\alpha \beta} )\;
\Psi_{\beta} \; ] ^{+}   =
\Psi^{+}_{\beta} \gamma^{0} \; ( \sqrt{3} g^{\alpha \beta} -
2 \sigma^{\alpha \beta} ) \; \Psi \; .
$$

\noindent Next, inserting after  $\Psi^{+}_{\beta} \gamma^{0}$ the
product $\eta  (x) \eta (x) ^{-1}$, get to the relation we need
$$
[ \;\bar{\Psi} \; ( \sqrt{3} g^{\alpha \beta} + 2 \sigma^{\alpha \beta} ) \;
\Psi_{\beta}\; ] ^{+}   =
\bar{\Psi} _{\beta}  \; ( \sqrt{3} g^{\beta \alpha } -
2 \sigma^{\beta \alpha } ) \; \Psi \; .
\eqno(4.19)
$$

\noindent  Thus, the current defined by (4.18) is a real-valued generally  covariant vector.

Else one question should be clarified. What will the current look  like in absence of  any external
electromagnetic field. Obviously,  an  expected result  is that the generalized current must
coincide with the conventional Dirac current.
To this end in view, let us compare expressions for  generalized
and conventional Dirac currents. Having used eqs.  (3.1),(3.2) and conjugate eqs.  (3.11a,b),
for the current according to  (3.18) one can easily find  the following expression
$$
 J^{\alpha} (x) = \bar{\Psi} \gamma^{\alpha}(x) \Psi - { 4\mu \over M}\; [ \;
\bar{\Psi} \sigma^{\alpha \beta} (x)  (\stackrel{\rightarrow}{\partial}_{\beta} + B_{\beta}) \Psi    + \bar{\Psi}
( \stackrel{\leftarrow}{\partial} _{\beta} - B_{\beta} ) \sigma^{\alpha \beta}(x) \Psi \; ] \;
,
\eqno(4.20a)
$$

\noindent that can be brought to the form
$$
J^{\alpha} (x) = \bar{\Psi} \gamma^{\alpha}(x) \Psi \; - \;
{4 \mu \over M}\;  \nabla_{\beta} \;
[ \; \bar{\Psi} \sigma^{\alpha \beta} (x)  \;   \Psi \; ] \; .
\eqno(4.20b)
$$

\noindent Therefore, in absence of external electromagnetic
fields, the above generalized conserved current differs from the
conventional current of  the Dirac particle only in the
term proportional to $\nabla_{\alpha} \Omega^{[\alpha \beta]}(x)$.
But such a freedom in the choice of any current is inherent in the theory.
In view of symmetry of the Ricci tensor, presented in  (3.20b) additional $\mu$-term  will  vanish identically in the
conservation law; indeed,
$$
\nabla_{\alpha }  \nabla_{\beta}
[ \; \bar{\Psi} \sigma^{\alpha \beta} (x)   \; \Psi \; ] = R_{\alpha \beta} \;
[ \; \bar{\Psi} \sigma^{\alpha \beta} (x) \;   \Psi \; ]  \equiv 0 \; .
$$

\subsection*{5. Equation for a main bispinor $\Psi$ }

In Section 5 we are going to study description of the generalized fermion in the representation
when an auxiliary vector-bispinor component is excluded from the equations. To this end,
let us turn again to the equations (4.1) and  (4.2):
$$
\gamma^{\alpha} D_{\alpha} \Psi  + \mu \;
(\; \sqrt{3} g^{\alpha \beta}  + 2 \sigma^{\alpha \beta}  \; )\;
     D_{\alpha} \Psi _{\beta}  - M \; \Psi  = 0 \; ,
\eqno(5.1)
$$
$$
\mu \;
(\;\sqrt{3} \delta_{\beta}^{\;\;\rho} -
 2 \sigma_{\beta}^{\;\;\rho} \; ) \;  D_{\rho} \Psi  - M \; \Psi_{\beta}  = 0 \; .
\eqno(5.2)
$$

\noindent
Substituting $\Psi_{\beta}$ from  (5.2) in (5.1), we get to
$$
\gamma^{\alpha} D_{\alpha} \Psi  + {\mu^{2} \over M} \;
 ( \sqrt{3} \;g^{\alpha \beta}  + 2 \sigma^{\alpha \beta}  )\;
(\sqrt{3} \; \delta_{\beta}^{\;\;\rho} -
 2 \sigma_{\beta}^{\;\;\rho} )  \; D_{\alpha}   D_{\rho} \Psi
 - M \; \Psi  = 0 \; ,
$$

\noindent    and further (see  (1.6))
$$
\gamma^{\alpha} D_{\alpha} \Psi  -  {4\mu^{2} \over M} \;
\sigma^{\alpha \beta } \;  D_{\alpha} D_{\beta} \Psi
 - M \Psi  = 0 \; .
\eqno(5.3)
$$

The commutator $[ D_{\alpha} , D_{\beta} ]_{-} $ is to be considered more closely:
$$
[ D_{\alpha} , D_{\beta} ]_{-}  \Psi = (D_{\alpha} D_{\beta} - D_{\beta} D_{\alpha}) \;\Psi =
$$
$$
= [\;  -ig F_{\alpha \beta}  +  (\partial_{\alpha}B_{\beta} -
\partial_{\beta}B_{\alpha} ) +
(B_{\alpha} B_{\beta} - B_{\beta} B_{\alpha}) \;  ] \;  \Psi \; ,
\eqno(5.4)
$$

\noindent
where $ D_{\alpha} = \nabla_{\alpha} + B_{\alpha} - ig A_{\alpha}$.
Take notice on the fact that on the right in (5.4) are present only algebraic expressions
(with no derivative over $\Psi$). The second term on the right
reads in more  detailed form as
$$
\partial_{\alpha } B_{\beta}  -
\partial_{\beta }  B_{\alpha} = \nabla_{\alpha} B_{\beta} -  \nabla_{\beta} B_{\alpha} =
$$
$$
= {1 \over 2} \sigma^{ab} \nabla_{\alpha} \;
 ( \; e_{(a)}^{\nu} e_{(b)\nu ; \beta} \;) \; -   \;
  {1 \over 2} \sigma^{ab} \nabla_{\beta}  \;
( \;e_{(a)}^{\nu} e_{(b)\nu ; \alpha} \; ) =
$$
$$
= {1 \over 2} \sigma^{ab} e_{(a)}^{\nu} \; [\; e_{(b)\nu ; \beta ; \alpha } -
                                      e_{(b)\nu ; \alpha ; \beta }\; ] \; +
{1 \over 2 } \sigma^{ab} \; [\; e_{(a) \nu ; \alpha } e_{(b) ; \beta}^{\nu} -
                      e_{(a) \nu ; \beta} e_{(b) ; \alpha}^{\nu} \; ] \; .
\eqno(5.5)
$$

\noindent
Similarly, the third is
$$
 B_{\alpha} B_{\beta} - B_{\beta} B_{\alpha}   =
$$
$$
=
(\; {1 \over 2} \sigma^{ab} e_{(a)}^{\nu} e_{(b)\nu ; \alpha }  ) \;
(\; {1 \over 2} \sigma^{kl} e_{(k)}^{\mu} e_{(l)\mu ; \beta } \; ) \; - \;
(\; {1 \over 2} \sigma^{kl} e_{(k)}^{\mu} e_{(l)\mu ; \beta }  ) \;
(\; {1 \over 2} \sigma^{ab} e_{(a)}^{\nu} e_{(b)\nu ; \alpha } \; ) \; =
$$
$$
= {1 \over 4} \left ( \sigma^{ab} \sigma^{kl}    -  \sigma^{kl}  \sigma^{ab} \right  ) \left [
\;   (  e_{(a)}^{\nu} e_{(b)\nu ; \alpha }  ) \;
\; e_{(k)}^{\mu} e_{(l)\mu ; \beta } \; )  \right ] \; ,
$$

\noindent that with the use of the known commutation relations
$$
[\sigma^{ab},\sigma^{kl}]_{-} =
( -\sigma^{kb} g^{la} + \sigma^{lb} g^{ka} ) - (  - \sigma^{ka} g^{lb}  +
\sigma^{la} g^{kb} ) \; ,
$$

\noindent leads to
$$
( B_{\alpha} B_{\beta} - B_{\beta} B_{\alpha}  ) =
 - {1 \over 2 } \sigma^{ab} \; [\; e_{(a) \nu ; \alpha } e_{(b) ; \beta}^{\nu} -
                      e_{(a) \nu ; \beta} e_{(b); \alpha}^{\nu} \; ] \; .
\eqno(5.6)
$$

\noindent Summarizing (5.5) and (5.6), we arrive at
$$
[D_{\alpha} D_{\beta}]_{-} =
  -ig F_{\alpha \beta} +
 {1 \over 2} \sigma^{ab}\; e_{(a)}^{\nu} \;
[\; e_{(b)\nu ; \beta ; \alpha } -
e_{(b)\nu ; \alpha ; \beta }\; ] =
$$
$$
=
-ig F_{\alpha \beta}+
{1 \over 2} \sigma^{ab}\; e_{(a)}^{\nu} \;
[ \;e_{(b)} ^{\rho} R_{\rho \nu \beta \alpha} \; ] =
- i g  F_{\alpha \beta} +
{1 \over 2 } \sigma^{\nu \rho} \; R_{\nu \rho \alpha \beta } \; ,
\eqno(5.7)
$$

\noindent here  $R_{\nu \rho \alpha \beta }(x) $ designates the Riemann curvature tensor.
Allowing for (5.7), the above  equation  (5.3) reads as
$$
[\; \gamma^{\alpha} D_{\alpha}   -  {2\mu^{2} \over M} \;
(\;  - i g \; \sigma^{\alpha \beta} F_{\alpha \beta} +
{1 \over 2 } \sigma^{\alpha \beta } \sigma^{\nu \rho} \;
R_{\nu \rho \alpha \beta }\; )\;
    - M \; ] \; \Psi  = 0 \; ,
\eqno(5.8)
$$

No we are to look more closely to the term
$ \sigma^{\alpha \beta } \sigma^{\nu \rho} R_{\nu \rho \alpha \beta }$.
In the first place, taking in mind the symmetry properties of the curvature tensor,
it can be rewritten in the form
$$
\sigma^{\alpha \beta } \sigma^{\nu \rho} \;
R_{\nu \rho \alpha \beta } = {1 \over 4}   \;  \gamma^{\alpha} \gamma^{\beta}
\gamma^{\nu} \gamma^{\rho} \;  R_{\nu \rho \alpha \beta } \; .
$$

\noindent Next, with the use of the  extended multiplication law for Dirac matrices, which can be
produced from the known formula  (3.2) in  the case of Minkowski space by multiplying it with
three tetrads)
$$
\gamma^{\alpha} \gamma^{\beta} \gamma^{\nu } =
\gamma^{\alpha} g^{\beta \nu } - \gamma^{\beta } g^{\alpha \nu } +
 \gamma^{\nu} g^{\alpha \beta} +
i \gamma^{5}
\epsilon^{\alpha \beta \nu \delta } \; \gamma_{\delta } \; ,
\eqno(5.9)
$$

\noindent
and taking into account that in view of the known symmetry property of the curvature tensor
with respect to cyclic permutation over any three indices its convolution with Levi-Civita tensor over three
indices  $\epsilon^{\alpha \beta \rho \sigma} R_{\alpha \beta \rho \delta}$
vanishes identically, we arrive at
$$
\sigma^{\alpha \beta } \sigma^{\nu \rho} \;
R_{\nu \rho \alpha \beta } = {1 \over 4} \;
(\gamma^{\alpha} g^{\beta \nu } - \gamma^{\beta } g^{\alpha \nu } )
\gamma^{\rho} \;  R_{\nu \rho \alpha \beta } \; ,
$$

\noindent or
$$
\sigma^{\alpha \beta } \sigma^{\nu \rho}
R_{\nu \rho \alpha \beta } = - {1 \over 2}
\gamma^{\alpha} \gamma^{\rho} \; R_{\alpha \rho}  = -{1 \over 2} \; R  \;
,
\eqno(5.10)
$$

\noindent where $R(x)$ stands for the Ricci scalar curvature.
Substituting (5.10) in   (5.8),  the latter  takes on the form
$$
\{ \; \gamma^{\alpha} D_{\alpha}  -  {2\mu^{2} \over M} \;
[ - i g  \sigma^{\alpha \beta} F_{\alpha \beta} -
{1 \over 4 }  R(x) ]   - M \;  \} \; \Psi  = 0 \; ,
\eqno(5.11)
$$

In absence of electromagnetic fields, eq. (5.11) read as  a modified Dirac equation
$$
[\;  \gamma^{\alpha} (\partial_{\alpha} + B_{\alpha} )   -   M +  {\mu^{2} \over 2M} \;
 R(x) \; ]\;  \Psi  = 0 \; ,
\eqno(5.12)
$$

\noindent or in ordinary dimensional units
$$
[\;i \;  \gamma^{\alpha} (\partial_{\alpha} + B_{\alpha} ) -
(mc/  \hbar ) \;  +  \; \mu^{2}
{ R (x) \over 2 (mc / \hbar)}  \; ]\;  \Psi  = 0 \; .
\eqno(5.13)
$$

\noindent
Take notice that eq. (5.13)  is not a common Dirac equation but involving
one additional $R$-term.
So, 20-component theory of a fermion particle implies both an
anomalous magnetic momentum  and additional gravitational
interaction though scalar Ricci $R$. And what is the most
interesting, the same single parameter $\mu$ influences both interaction terms as an intensity
multiplier.

\subsection*{6. Masless limit and conformal invariance}

In massless case, instead of eqs.  (5.1) and (5.2), one is to employ the following
equations\footnote{To avoid misunderstanding a few comments should be done.
From heuristic considerations, in massless case one has at least a theoretical possibility
(bearing in mind the gauge invariance principle) to investigate a massless complex-valued field in external
vector ("electromagnetic") field. At this, the term "electromagnetic" must be understood with caution,
in fact as a matter of convention.}
$$
\gamma^{\alpha} D_{\alpha} \Psi  + \mu \;
 (  \sqrt{3} \; g^{\alpha \beta}  + 2 \sigma^{\alpha \beta} )\;
          D_{\alpha} \Psi _{\beta}   = 0 \; ,
\eqno(6.1)
$$
$$
\mu \;
(\sqrt{3} \; \delta_{\beta}^{\;\;\rho} -
 2 \sigma_{\beta}^{\;\;\rho} 0 \; D_{\rho} \Psi  =  \Psi_{\beta}   \; .
\eqno(6.2)
$$

\noindent  These equations, after exclusion of the auxiliary vector-bispinor
component, result in
$$
[ \gamma^{\alpha} D_{\alpha}   -  2\mu^{2}  \;
( - i g  \sigma^{\alpha \beta} F_{\alpha \beta} -
{1 \over 4 }  R ) \; ] \; \Psi  = 0 \; ,
\eqno(6.3)
$$

\noindent
If the electromagnetic field vanishes, eq. (6.3)  reads as
$$
[\;  \gamma^{\alpha} (\partial_{\alpha} + B_{\alpha} )   +
 {1 \over 2} \mu^{2} R(x) \; ]\;  \Psi  = 0 \; .
\eqno(6.4)
$$

\noindent In contrast to  the massive case, here the $\mu$ represents yet another (in particular, dimensional
one)  characteristic, which  is  absolutely different from the old one in the  massive case.

Now we are to deviate from the main line our consideration above and  are going to study  in some
detail one theoretical criterion for correctness of any massless wave equation as applied to the
20-component fermion.
As previous experience has shown we may expect that any wave equations for massless particles be
invariant under conformal transformations; so is
electromagnetic field, so is a massless Dirac equation, and so is
a conform-invariant equation for a scalar particle.

With this end in view, let us introduce some  conventions and formulas.
Ordinary Dirac massless equation in a space-time with metric $g_{\alpha \beta}$ has the form
$$
[ \; i \; \gamma ^{\alpha }(x)\; (\partial_{\alpha} \;  +  \;
B_{\alpha } ) \;\Phi  (x)  = 0 \; .
\eqno(6.5)
$$

\noindent Let metric tensors of the two space-time models differ in an arbitrary factor-function
as shown below
$$
dS^{2} = g_{\alpha \beta} dx^{\alpha} dx^{\beta} \; , \;
d\tilde{S}^{2} = \tilde{g}_{\alpha \beta} dx^{\alpha} dx^{\beta} \; , \;
$$
$$
\tilde{g}_{\alpha \beta}(x) = \varphi^{2} (x) \; g_{\alpha \beta} (x) \; .
\eqno(6.6)
$$

\noindent
Correspondingly, it is the most simple to choose   tetrads proportional to each other
$$
\tilde{e}_{(a)}^{\alpha} = {1 \over  \varphi } e_{(a)}^{\alpha}  \; , \qquad
\tilde{e}_{(a)\alpha} =   \varphi \;  e_{(a)\alpha}  \; .
\eqno(6.7)
$$

Besides, we need to compare two sets of Christoffel symbols. Starting from the relation below
$$
\tilde{\Gamma}_{\alpha \beta, \rho} (x) =
{1 \over 2} \; [\;
\partial_{\alpha} \tilde{g}_{\beta \rho} + \partial_{\beta} \tilde{g}_{\alpha \rho}
- \partial_{\rho} \tilde{g}_{\alpha \beta} \; ] =
{1 \over 2} \; [ \;
\partial_{\alpha} \varphi^{2} g_{\beta \rho} +
\partial_{\beta} \varphi^{2} g_{\alpha \rho}
- \partial_{\rho} \varphi^{2} g_{\alpha \beta}  ) =
$$
$$
= \varphi^{2} \Gamma_{\alpha \beta, \rho} + \varphi \; [
(\partial_{\alpha} \varphi ) g_{\beta \rho} +
(\partial_{\beta} \varphi ) g_{\alpha \rho}
- (\partial_{\rho} \varphi ) g_{\alpha \beta} \; ] \; ,
$$

\noindent or
$$
\tilde{\Gamma}^{\sigma}_{\alpha \beta}(x) =
\tilde{g}^{\sigma \rho}
\Gamma_{\alpha \beta, \rho} = { 1 \over \varphi^{2}} g^{\sigma \rho}
\{ \varphi^{2} \Gamma_{\alpha \beta, \rho} + \varphi \;
[(\partial_{\alpha} \varphi ) g_{\beta \rho} +
(\partial_{\beta} \varphi ) g_{\alpha \rho}
- (\partial_{\rho} \varphi ) g_{\alpha \beta} \; ] \} \; .
$$

\noindent we obtain
$$
\tilde{\Gamma}^{\sigma}_{\alpha \beta}(x) =
\Gamma^{\sigma}_{\alpha \beta}(x) + { 1 \over \varphi} \;
[ \; (\partial_{\alpha} \varphi ) \delta^{\sigma}_{\beta}  +
(\partial_{\beta} \varphi ) \delta^{\sigma}_{\alpha}
- g^{\sigma \rho }(x)  (\partial_{\rho} \varphi ) g_{\alpha \beta} \; ]  \;  \; .
\eqno(6.8)
$$

Similarly, for two sets of Ricci rotation coefficients we have
$$
\tilde{\gamma}_{abc}(x) =
-   \; [ \; {\partial \over x^{\beta}} \tilde{e}_{(a)\alpha}   -
\tilde{\Gamma}^{\rho} _{\beta \alpha} \tilde{e}_{(a)\rho} \; ] \;
\tilde{e}_{(b)}^{\alpha} \tilde{e}_{(c)}^{\beta} =
$$
$$
=  -  [ \;
{\partial \over \partial x^{\alpha} }  ( \varphi e_{(a)\alpha} )-
\Gamma^{\rho}_{\beta \alpha} \; \varphi e_{(a)\rho} -
$$
$$
-
{1 \over \varphi }  \;
\left ( \;(\partial_{\beta} \varphi ) \delta^{\rho}_{\alpha}  +
(\partial_{\alpha} \varphi ) \delta^{\rho}_{\beta}
- g^{\rho \sigma}(x)  (\partial_{\sigma} \varphi ) g_{\beta \alpha }(x) \right )\;
\varphi e_{(a)\rho } \;
\; ] \;   {1 \over \varphi^{2} } e_{(b)}^{\alpha} e_{(c)}^{\beta}  \; .
$$

\noindent a further
$$
\tilde{\gamma}_{abc}(x) = -
[ \; \varphi \;  e_{(a)\alpha; \beta}  - {\partial  \varphi \over \partial x ^{\alpha} }
e_{(a)\beta} + e_{(a)}^{\sigma}
{ \partial   \varphi \over \partial x^{\sigma} } g_{\alpha \beta} (x) \;  ]
 \; {1 \over \varphi^{2} } e_{(b)}^{\alpha} e_{(c)}^{\beta} \; .
$$

\noindent The final formula is
$$
\tilde{\gamma}_{abc}(x) =
 { 1 \over \varphi } \; \gamma_{abc}(x)  + { 1 \over \varphi^{2}} \;
[\; e_{(b)}^{\sigma}   g_{ac}
 \partial_{\sigma}  \varphi    -
e_{(a)}^{\sigma}  \; g_{bc} \partial_{\sigma}  \varphi \; ] \;.
\eqno(6.9)
$$

Now, we are ready to relate two Dirac equations associated with different space-time models.
Let us  start with that in  $\tilde{g}_{\alpha \beta}(x)$-space:
$$
 i \; \tilde{\gamma} ^{\alpha }(x)\; (\partial_{\alpha} \;  +  \;
\tilde{B}_{\alpha }(x) ) \; \tilde{\Phi}  (x)  = 0 \; ,
\eqno(6.10)
$$

\noindent or in altered form
$$
 i \gamma^{c} \; ( \tilde{e}_{(c)}^{\alpha} \partial_{\alpha} +
{1 \over 2} \sigma^{ab} \tilde{\gamma}_{abc} )  \; \tilde{\Phi} = 0 \; .
\eqno(6.11)
$$

Substituting here the above expressions of the 'tilded' tetrad and Ricci rotation coefficients in terms of
'untilded' ones we arrive at
$$
\{ \;[\; i \gamma^{c} \; ( e_{(c)}^{\sigma} \partial_{\sigma} +
{1 \over 2}  \sigma^{ab} \gamma_{abc} ) \;+
$$
$$
+ { i \over 2} \; \gamma ^{c} \sigma^{ab}
\;[ \; ( e_{(b)}^{\sigma}   g_{ac}
 -  e_{(a)}^{\sigma}  \; g_{bc} )\;
{1 \over \varphi }
  \partial_{\sigma}  \varphi \; ] \; \}\; \tilde{\Phi} = 0 \;,
\eqno(6.12)
$$

\noindent which is equivalent to
$$
[\; i \gamma^{\sigma} \; ( \partial_{\sigma} + B_{\sigma})
+   i  \; \gamma _{a} \sigma^{ab} e_{(b)}^{\sigma} \;
  {1 \over \varphi } \partial_{\sigma} \varphi \; ]\; \tilde{\Phi} = 0 \;.
\eqno(6.13)
$$

\noindent
Because
$$
\gamma_{a} \; \sigma ^{ab} = {3 \over 2} \gamma^{b} \; ,
\eqno(6.14)
$$

\noindent eq. (6.13)  will read as
$$
i \gamma^{\sigma}(x) \; [ \; \partial_{\sigma} + B_{\sigma} (x)
+ (3 / 2) \;  { 1 \over \varphi }  \;
  \partial_{\sigma}  \varphi \; ] \; \tilde{\Phi} = 0 \; .
\eqno(6.15)
$$

\noindent
And the final step is to have used the following  substitution
$$
\tilde{\Phi}(x) = (\varphi^{-1})^{-3/2} \Phi (x)
\eqno(6.16)
$$

\noindent at this eq. (6.15) describing a massless fermion  field $\tilde{\Phi}(x)$ in the space-time
with $\tilde{g}_{\alpha \beta}(x)$-metric will take on the form of Dirac massless wave
equation in  the 'tilded' space-time:
$$
i \gamma^{\sigma}(x) [  \partial_{\sigma} + B_{\sigma} (x)
 \; ] \; \Phi = 0 \; ,
\eqno(6.17)
$$

\noindent
This fact of such a simple relationship between two Dirac equations in $g$- and $\tilde{g}$-model is
referred to as the conformal invariance  property of the massless Dirac equation [62].

Now, in a similar manner, let us consider  the case of a particle with anomalous magnetic momentum.
We will need one auxiliary formula . The Ricci scalar behaves
under conformal  transformations according to  the law (G\"{u}rsey F. Ann. Phys.  1963. Vol. 24, P.
 211-244;
also see Appendix A)
$$
\tilde{R} =
{1 \over \varphi^{2}} \;  (\;
R -  {6 \over \varphi }\; \nabla^{\beta} \nabla_{\beta} \varphi  )\; .
\eqno(6.18)
$$

\noindent On taking into account this law we immediately have to conclude that eq. (6.4) is not
conformally invariant.

Obviously, we might start with a bit different equations (compare with (6.1) and (6.2)):
\footnote{The same $R$-term might be added in the massive case too.}:
$$
\gamma^{\alpha} D_{\alpha} \Psi  + \mu \;
 [\; \sqrt{3} \; g^{\alpha \beta}  + 2 \;\sigma^{\alpha \beta} \; ]\;
          D_{\alpha} \Psi _{\beta}   =  {1 \over 2} \mu^{2}\; R(x) \; \Psi  \; ,
\eqno(6.19)
$$
$$
\mu\;
(\;\sqrt{3} \delta_{\beta}^{\;\;\rho} -
 2 \sigma_{\beta}^{\;\;\rho} \; ) \; D_{\rho} \Psi  =  \Psi_{\beta}   \; .
\eqno(6.20)
$$

\noindent Consequently, after exclusion the vector-bispinor constituent from eqs. (6.19) it follows
$$
[\; \gamma^{\alpha}(x) \; (\partial_{\alpha} +
B_{\alpha} - i {e \over \hbar c } A_{\alpha} )   +  2 i \mu^{2}  {e \over \hbar c}  \;
 \sigma^{\alpha \beta} (x) \;  F_{\alpha \beta}(x) \; ] \; \Psi  = 0 \; ,
\eqno(6.21)
$$

\noindent that is  a massless conformally invariant Dirac equation.
In absence of an external 'electro\-magnetic' field it yields
$$
\gamma^{\alpha}(x)  (\partial_{\alpha} + B_{\alpha} ) \; \Psi = 0 \; .
\eqno(6.22)
$$

\subsection*{7. On gauge $P$-symmetry of the theory}

In this Section we will consider some features concerned with a tetrad $P$-symmetry of the 20-component
fermion model.  In the first place, several initial fact and notations are to be recalled
необходимые обозначения).

As known, the ordinary general covariant Dirac equation
$$
[ \; i \gamma^{\alpha} (x) (\partial_{\alpha} + B_{\alpha}) - m \; ] \; \Psi (x) = 0
\eqno(7.1)
$$

\noindent possesses a symmetry under the following discrete transformation
(we  assume the use of the   Weyl spinor  basis)
$$
\Psi ' (x) = \gamma^{0} \; \Psi (x) \; .
\eqno(7.2)
$$

\noindent
Let us demonstrate this. First, we need  two relationships
$
\gamma^{0} \gamma^{0} \gamma^{0} =  + \gamma^{0}  \; , \;
\gamma^{0} \gamma^{i} \gamma^{0} =  - \gamma^{i}  \; , \;
$
hold, which  can be summarized  in the formula
$$
\gamma^{0} \gamma^{a} \gamma^{0}  = \bar{\delta}^{a}_{b}  \gamma^{b} =
  \gamma^{\;b}  \; L^{(p)\; a}_{b}  \;
\eqno(7.3)
$$

\noindent with the use of a  special designation for Lorentz $P$-inversion:
$$
 L^{(p)\; a}_{b}   =\bar{\delta}^{a}_{b}  =
\left ( \begin{array}{cccc}
+1 &  0  &  0  &  0  \\
0  & -1  &  0  &  0  \\
0  &  0  &  -1 &  0  \\
0  &  0  &  0  & -1
\end{array} \right ) \; .
$$

Further, taking into account eq.  (7.3), we produce the transformation laws for ther generalized Dirac matrices
and bispinor connection:
$$
\gamma^{0} \gamma^{\alpha}(x) \gamma^{0} = \gamma^{'\alpha} (x) \; , \;
\gamma^{0} B_{\alpha}(x) \gamma^{0} = B'_{\alpha} (x) \; , \;
\eqno(7.4a)
$$

\noindent  where primed quantities are built  with the use of  the (primed) $P$-inverted tetrad
$$
e'_{(b) \alpha} (x)  = L^{(p)\; a }_{b}  e_{(a) \alpha} (x) \; .
\eqno(7.4b)
$$

\noindent With eqs. (7.4),  from (7.2) it follows an equation for $\Psi' (x)$:
$$
[\; i \gamma^{'\alpha} (x) \; (\partial_{\alpha} + B'_{\alpha}) \;-\; m \; ] \; \Psi '(x) = 0 \; ,
\eqno(7.5)
$$

\noindent which formally coincides with eq. (7.1). This fact proves the tetrad  $P$-symmetry we need.

Now let us turn to 20-component model. By simplicity reasons, it is convenient firstly  to study
the case of a flat space-time ( that is $D_{a} = \partial_{a} - igA_{a}(x)$)
$$
\gamma^{a} \; D_{a} \Psi  + \mu \; Z^{ab} \; D_{a} \; \Psi_{b} = M \;\Psi  \; ,
\eqno(7.6a)
$$
$$
\mu \; Y_{b}^{\;\;a} \; D_{a} \Psi = M \;\Psi_{b} \; ,
\eqno(7.6b)
$$

\noindent where
$$
Z^{ab} =  (\sqrt{3} \;g^{ab} + 2 \sigma^{ab}) \; , \;\;
Y_{a}^{\;\;b} = ( \sqrt{3} \;\delta^{\;\;b}_{a} - 2 \sigma_{a}^{\;\;b} ) \; .
$$

\noindent As concerns to these notations, two useful equalities should be noted:
$$
Z^{ab} = Y^{ba} \; , \qquad \gamma^{0} \; (Z^{ab})^{+} \gamma^{0} =  Y^{ab} = Z^{ba} \;.
$$

\noindent The $P$-inversion here as applied to 20-component field is defined by
$$
x'_{b} = L_{b}^{(p) \; a}\;  x_{a} \; , \qquad
\Psi ' (x') = \gamma^{0}\;  \Psi (x) \; , \qquad
\Psi' _{b} (x') =   \gamma^{0} \;  L_{b}^{(p) \; a}  \; \Psi _{b}(x) \; .
\eqno(7.7)
$$

\noindent With the use of eq.  (7.3) and of  these
$$
\gamma^{0}\; Z^{ab} \; \gamma^{0} = \bar{\delta}^{a}_{k} \;
\bar{\delta}^{b}_{l}\;  Z^{kl}\; ,\qquad
\gamma^{0} \; Y^{\;\;a}_{b}\; \gamma^{0} = \bar{\delta}^{k}_{b} \; \bar{\delta}^{a}_{l} \;
Y_{k}^{\;\;l} \; ,
\eqno(7.8)
$$

\noindent from (7.6) one readily gets the primed equations
$$
\gamma^{a} \; D'_{a} \Psi'  + \mu \; Z^{ab} \; D'_{a} \; \Psi'_{b} = M \; \Psi' (x) \; ,
\eqno(7.9a)
$$
$$
\mu  \; Y_{b}^{\;\;a} \; D'_{a} \; \Psi' = M \; \Psi'_{b} \; ,
\eqno(7.9b)
$$

\noindent where $D'_{a} = L_{a}^{(p)\; b} D_{b}$.
So,  20-component fermion model in Minkowski space-time is $P$-invariant.

In will be  useful  to have in hand a formulation of this  symmetry  in two-component spinor basis,
which can be derived quite straightforwardly. Here eqs. (7.1) will take the form
(see. (2.4) и (2.5))
$$
\bar{\sigma}^{a}  D_{b}  \;\eta +
\mu \; z^{ab} D_{a} \; \xi _{b} = M \; \xi \; ,
\eqno(7.10a)
$$
$$
\sigma^{a} D_{\alpha} \;  \xi  +   \mu \; \bar{z}^{ab} D_{a} \; \eta _{b} =
 M \; \eta  \; ,
\eqno(7.10b)
$$
$$
\mu  \; y_{b}^{\;\;c}  \;   D _{c} \; \xi = M  \; \xi_{b} \; ,
\eqno(7.11a)
$$
$$
\mu \;  \bar{y}_{b}^{\;\;c} D_{c} \;  \eta = M \; \eta_{b} \; .
\eqno(7.11b)
$$

\noindent
The notation is as follows
$$
z^{ab} =  (\sqrt{3}\; g^{ab} + 2 \Sigma^{ab}) \; , \;\;
\bar{z}^{ab} =  (\sqrt{3}\; g^{ab} + 2 \bar{\Sigma} ^{ab}) \; , \;\;
$$
$$
y_{a}^{\;\;b} = ( \sqrt{3} \; \delta^{\;\;b}_{a} - 2 \Sigma_{a}^{\;\;b} ) \; ,  \;\;
\bar{y}_{a}^{\;\;b} = ( \sqrt{3} \; \delta^{\;\;b}_{a} - 2 \bar{\Sigma}_{a}^{\;\;b} ) \; ,
\eqno(7.12)
$$

\noindent and relations $
z^{ab} = y^{ba} \; ,  \; \;  \bar{z}^{ab} = \bar{y}^{ba}  \; $ hold.
$P$-transformation in spinor form looks as (see  (7.7))
$$
\xi ' (x')   = + \; \eta (x) \; , \; \; \eta '(x') = + \; \xi (x) \;
$$
$$
\xi '_{b}  (x')   = +\; \bar{\delta}_{b}^{a} \;   \eta_{a} (x) \; , \; \;
\eta '_{b} (x') = + \;  \bar{\delta}_{b}^{a} \;  \xi _{a} (x) \;
\eqno(7.13)
$$

\noindent
Substituting in (7.10) and (7.11) the components
 $\xi , \eta ,  \xi_{b} , \eta_{b} $ in terms of the
 $\xi' , \eta ',  \xi'_{b} , \eta '_{b}$ according to (7.12),
and bearing in mind the formulas
$$
\bar{\sigma}^{a} = \bar{\delta}^{a}_{b} \; \sigma^{b} \; , \;\;
\sigma^{a}  = \bar{\delta}^{a}_{b} \; \bar{\sigma}^{b} \; , \;\;
$$
$$
\bar{z}^{ab} = \bar{\delta} ^{a}_{k}  \;     \bar{\delta} ^{b}_{l} \; z^{kl} \;  , \;\;
z^{ab} = \bar{\delta} ^{a}_{k} \; \bar{\delta} ^{b}_{l} \; \bar{z}^{kl} \;  , \;\;
$$
$$
\bar{y} _{b}^{\;\;c}  = \bar{\delta}_{b}^{k}  \; \bar{\delta}^{c}_{l}  \;
y_{k} ^{\;\;l} \; , \;\;
y _{b}^{\;\;c}  = \bar{\delta}_{b}^{k} \;  \bar{\delta}^{c}_{l}  \;
\bar{y}_{k} ^{\;\;l} \; , \;\;
\eqno(7.14)
$$

\noindent from  (7.10) and (7.11) $P$-inverted equations follow
$$
\bar{\sigma}^{a} D'_{b}  \; \eta' +
\mu \; z^{ab} D'_{a} \; \xi' _{b} = M \; \xi' \; ,
\eqno(7.15a)
$$
$$
\sigma^{a} D'_{\alpha} \;  \xi'  +   \mu \; \bar{z}^{ab} D'_{a} \; \eta' _{b} =
 M \; \eta'  \; ,
\eqno(7.15b)
$$
$$
\mu  \; y_{b}^{\;\;c}    D' _{c}  \; \xi' = M \; \xi'_{b} \; ,
\eqno(7.16a)
$$
$$
\mu \;  \bar{y}_{b}^{\;\;c} D'_{c} \;  \eta' = M \; \eta'_{b} \; ,
\eqno(7.16b)
$$

\noindent which formally coincide with those  original (7.10), (7.11).

At last, we  are to look at the generally covariant 20-component case. The situation in the chosen basis
$\Psi (x), \Psi _{\beta}(x)$ turns out to be much simpler than in Minkowski space-time because
the equations under consideration
$$
\gamma^{\alpha} D_{\alpha} \; \Psi  + \mu \;  Z^{\alpha \beta }
D_{\alpha} \; \Psi_{\beta} = M \; \Psi  \; ,
\eqno(7.17a)
$$
$$
\mu \; Y_{\beta }^{\;\; \alpha } D_{\alpha } \; \Psi = M \; \Psi_{\beta} \; ;
\eqno(7.17b)
$$

\noindent are obviously form-invariant under the tetrad $P$-inversion established here in accordance
with the formulas (compare with  (7.2))
$$
\Psi ' (x) = \gamma^{0} \; \Psi (x) \; , \qquad
\Psi '_{\beta} (x) = \gamma^{0} \; \Psi_{\beta} (x) \; .
\eqno(7.18)
$$

\noindent Take notice that here the $P$-inversion operation over  the wave  function components
does not affect its  vector index, because it  is hidden in generally covariant vector  notation:
$\Psi_{\beta} = e^{(l)}_{\beta}(x) \Psi_{l} (x)$\footnote{Translating to the explicit use of a tetrad vector
index will be given in Sec. 8.}.

\subsection*{8. On matrix formulation of the   20-component theory}

Now, let us obtain a matrix form of the generalized fermion model\footnote{As will be seen below,
sometimes this formalism has  advantages over spin-vector approach.}.
To this end, we  turn again to eq. (2.1):
$$
\gamma^{\alpha} D_{\alpha} \; \Psi  + \mu  \;
 Z^{\alpha \beta}  D_{\alpha} \; \Psi _{\beta}  = M \; \Psi  \; ,
\eqno(8.1a)
$$
$$
\mu \; Y_{\beta}^{\;\;\rho} D_{\rho} \; \Psi  = M \; \Psi_{\beta}  \; .
\eqno(8.1b)
$$

\noindent and produce their form when the generally covariant vector index
of the wave function is translated into a tetrad one. We need a few auxiliary relations.
From
$$
\nabla_{\alpha} \;  \Psi_{\beta} =
\nabla_{\alpha} \; ( \Psi_{l} e^{(l)}_{\beta } ) =
\Psi_{l}\;  e^{(l)}_{\beta \;  ; \alpha}     +
e^{(l)}_{\beta} \partial_{\alpha} \;  \Psi_{l} \; ,
$$

\noindent it follows
$$
D_{\alpha}  \; \Psi_{\beta} =
e^{(l)}_{\beta} \partial_{\alpha} \;  \Psi_{l} +
e^{(l)}_{\beta \; ; \alpha} \; \Psi_{l}  +
B_{\alpha} \; e^{(l)}_{\beta}  \; \Psi_{l}    - i g A_{\alpha} \;  e^{(l)}_{\beta} \; \Psi_{l}  \;.
\eqno(8.2)
$$

\noindent
In turn, the vector connection according to Tetrode-Weyl-Fock-Ivanenko method
$$
L_{\alpha} = {1 \over 2} \; (V^{cd})_{k}^{\;\;l} \; e^{\beta}_{(c)} \nabla_{\alpha}
e_{(d)\beta}  \; .
\eqno(8.3a)
$$

\noindent by taking into account the explicit form of the generator for the vector
representation of the Lorentz group
$(V^{ab})_{k}^{\;\;l} =  ( -g^{al} \delta^{b}_{k}  + g^{bl} \delta_{k}^{a} ) \;$,
will read  as
$$
(L_{\alpha})_{k}^{\;\;l} =
e_{(k)}^{\beta} e^{(l)}_{\beta \; ; \alpha} \; = -
e_{(k)\; ; \alpha }^{\beta} \; e^{(l)}_{\beta \; }   \; .
\eqno(8.3b)
$$

\noindent
With the use of eqs. (8.2) and (8.3),  the equations  (8.1) yield
$$
e_{(a)}^{\alpha} \gamma^{a}   \;
(\partial_{\alpha} + B_{\alpha} - ig A_{\alpha}) \;  \Psi +
+  \mu \; Z^{ab} \;  e_{(a)}^{\alpha} \;  \times
$$
$$
\times \; [ \;
\delta_{b}^{\;\;l} \partial_{\alpha}  + (L_{\alpha})_{b}^{\;\;l} +
\delta _{b}^{\;\;l}  B_{\alpha} - ig \; \delta _{b}^{\;\;l}  A_{\alpha} \; ]
 \; \Psi_{l}  = M \; \Psi \; ,
\eqno(8.4a)
$$
$$
\mu \;  Y_{l}^{\;\;d} \;   e_{(d)}^{\; \alpha}
\; (\partial_{\alpha} + B_{\alpha}   - i g A_{\alpha} ) \; \Psi = M \; \Psi _{l} \; .
\eqno(8.4b)
$$

\noindent
Further, the  notation will be used:
$$
\Phi = \left ( \begin{array}{l} \Psi  \\ \Psi_{l}
\end{array} \right ) \; , \;
\Gamma^{a} =
\left (  \begin{array}{cc}
\gamma^{a}    &  \mu (Z^{a})_{(0)}^{\;\;\;\;b} \\
\mu (Y^{a})_{b}^{\;\;(0)}  &  0  \end{array}  \right ) =
\left (  \begin{array}{cc}
\gamma^{a}   & \mu Z^{ab} \\
\mu Z_{\;\;b}^{a}  &  0
\end{array} \right ) \; ,
$$
$$
J^{cd} = \sigma^{cd} \otimes I_{5} + I \otimes j^{cd} \; , \qquad j^{cd} =
\left ( \begin{array}{cc}  0  &  0  \\ 0  & (V^{cd})_{b}^{\;\;l} \end{array} \right ) \; ,
\eqno(8.5a)
$$
$$
\Gamma^{\alpha} (x) = \Gamma^{a} e_{(a)}^{\; \alpha}  \; , \qquad
G_{\alpha} = {1 \over 2} \; J^{cd} e_{(c)}^{\beta} \nabla_{\alpha} e_{(d)\beta} \; ,
$$
$$
G_{\alpha}(x) = \left ( \begin{array}{cc}
B_{\alpha}    & 0 \\ 0 &  B_{\alpha}
 \otimes I  + I \otimes L_{\alpha}
\end{array} \right ) \; .
\eqno(8.5b)
$$

\noindent
It is easily  verified that eqs. (8.4)  can be rewritten as the following generally  covariant
matrix equation:
$$
[ \; \Gamma^{\alpha}(x) ( \; \partial_{\alpha} + G_{\alpha} - ig A_{\alpha} \; )
- M \; ] \; \Phi = 0 \; .
\eqno(8.6)
$$

\noindent
Below, we will also exploit the notation
$$
D^{s.}_{\alpha} = \partial_{\alpha} + B_{\alpha} -ig A_{\alpha} \; , \qquad
D^{v.}_{\alpha} = \partial_{\alpha} + B_{\alpha} \otimes I_{4} + I \otimes L_{\alpha}
 -i g A_{\alpha} \; , \;
$$
$$
D_{\alpha} = \left ( \begin{array}{cc}
D^{s.}_{\alpha} &  0  \\
0  &   D^{v.}_{\alpha}       \end{array} \right )  \; ,
\eqno(8.7)
$$

\noindent with which  the equation (8.6) can be presented as
$$
[  \; \Gamma^{\alpha}(x)  \; D_{\alpha} - M \; ] \; \Phi = 0 \; .
\eqno(8.8)
$$

Now, we consider again the question about the Lorentz invariant form matrix in that approach.
Obvious, it suffices to restrict ourselves to the flat space-time case:
$$
\left ( \begin{array}{cc}    \gamma^{a}     &   \mu  Z^{al}   \\
\mu Y_{l}^{\;\; a}     &     0      \end{array}    \right )  \partial_{a}
\left ( \begin{array}{l}   \Psi  \\ \Psi_{l}  \end{array}  \right )  - M \;
\left ( \begin{array}{l}   \Psi  \\ \Psi_{l}  \end{array}  \right ) = 0  \;, \;
\;\; или \;\; (\Gamma^{a} \partial _{a} - M) \Phi = 0 \;  .
$$

\noindent
Following the usual procedure
$$
\bar{\Phi} ( \Gamma^{a} \stackrel{\leftarrow}{\partial}_{a} + M ) = 0 \; , \qquad
\bar{\Phi} = \Phi^{+} H \; ,
$$
$$
\Phi^{+} = ( \Psi^{+} , \Psi^{+}_{k} ) \; , \qquad
 H =  \left ( \begin{array}{cc}   \gamma^{0}  &  0  \\ 0  &  \gamma^{0} h^{kl}
\end{array} \right ) \; , \;
\eqno(8.9)
$$

\noindent we get to a defining relation for   $H$-matrix:
$$
H^{-1} (\Gamma^{a})^{+} H = \Gamma^{a} \; .
\eqno(8.10a)
$$

\noindent This equation, in turn, after taking into account the block-structure of the all matrices involved
leads to
$$
\left ( \begin{array}{cc}
\gamma^{0} \gamma^{a+} \gamma^{0}    &
 \mu \gamma^{0} Y^{a +}  \gamma^{0} h   \\
\mu  h^{-1} \gamma^{0} Z^{a+} \gamma^{0}  &  0  \end{array} \right ) =
\left ( \begin{array}{cc}
\gamma^{a}  & \mu Z^{a} \\
\mu Y^{a}  &  0     \end{array}  \right ) \; .
$$

\noindent So,
$$
\gamma^{0} \; \gamma^{a+} \; \gamma^{0}  = \gamma^{a} \; , \;\;\;
\gamma^{0} \; Y^{a+} \;\gamma^{0} \;  h = Z^{a}   \; , \; \;\;
h^{-1} \; \gamma^{0} \; Z^{a+}\; \gamma^{0} =  Y^{a} \; .
\eqno(8.10b)
$$

\noindent
The truth of the first of these has been already noted (see (4.4)).
Two remaining ones give respectively
$$
\gamma^{0} \; ( Z^{a}_{\;\;l} )^{+} \; \gamma^{0}\;  h^{lb}  = Z^{a b}   \; ,\qquad
(h^{-1})_{cl} \;  \gamma^{0} \; (Z^{al})^{+} \; \gamma^{0} =  Y^{\;\;a}_{c} \; .
$$

\noindent and further
$$
  Y^{a}_{\;\;\;l}  \;   h^{lb}  = Z^{a b}   \; , \qquad
(h^{-1})_{cl} \;   Z^{la}  =  Y^{\;\;\;a}_{c} \; .
\eqno(8.11)
$$

\noindent Here, the first one  represents already known relation (4.4a) being transformed to a tetrad basis,
whereas the second is a tetrad version  of the above equation (4.6b); at this an expression for $h$-matrix follows immediattely
in the form
$$
h^{lb} =  g^{lk} h^{\;\;b}_{k} \; , \qquad  h^{\;\;b}_{k} = \eta^{\;\;b}_{k} \;  .
\eqno(8.12)
$$

Thus, the conserved current expression (4.18) can be rewritten in a matrix formalism as follows
$$
J^{\alpha}  =   \bar{\Phi} \; \Gamma^{\alpha} \; \Phi =
\bar{\Psi} \; \gamma^{\alpha} \; \Psi  +
\mu \; ( \;  \bar{\Psi} \; Z^{\alpha \beta } \; \Psi_{\beta}   +
\bar{\Psi}_{\beta} \; Y^{\beta \alpha } \;  \Psi \; )  \; .
\eqno(8.13)
$$

Now let us turn again to a generally covariant equation (8.8) and  derive a conjugate one.
From  (8.8) it follows
$$
\bar{\Phi}  \;
[ \; ( \stackrel{\leftarrow}{\partial}_{\alpha} +
\; H^{-1} G_{\alpha}^{+} H \; + ig A_{\alpha} ) \; H^{-1} \Gamma^{\alpha +} H \; + \;
M \; ] = 0 \; .
\eqno(8.14)
$$

\noindent
Allowing for the equalities
$$
H^{-1} \gamma^{\alpha +} H  = + \gamma^{\alpha} \; , \qquad
H^{-1} G_{\alpha}^{+} H  = - \; G_{\alpha} \; ,
\eqno(8.15)
$$

\noindent from  (8.14) we get
$$
\bar{\Phi}  \;
[ \; ( \stackrel{\leftarrow}{\partial}_{\alpha} - \;
G_{\alpha} + ig A_{\alpha} ) \; \Gamma^{\alpha +} \; + \;
M \; ] = 0 \; .
\eqno(8.16)
$$

Evidently, the first relation in (8.15) is  a simple consequence from  (8.10a). The second in (8.15)
is easily verified on accounting for that  the $H$-matrix is to allow a Lorentz invariant to be of the form
$$
\bar{\Phi} ' H \Phi ' =  \bar{\Phi}  H \Phi  , \qquad   \Phi ' = S \; \Phi \; ,
$$

\noindent where $S$ represents a Lorentz transformation in 20-component $\Phi$-space. Therefore,
the following relationship
$$
H^{-1} \; S^{+} \; H = S^{-1} \;
\eqno(8.17)
$$

\noindent  is to  hold. From this, taking an infinitesimal transformation
$$
S = I + \delta_{ab} \; J^{ab} \; , \qquad
S^{-1} = I - \delta_{ab} \; J^{ab} \; , \qquad
S^{+}  = I + \delta_{ab} \; (J^{ab})^{+} \; , \;
$$

\noindent
we come to
$$
H^{-1} \; ( J^{ab})^{+}   H = - J^{ab} \; .
\eqno(8.18)
$$

\noindent The second relation in (8.15) is readily proved simply by allowing for the equation  (8.18) and
the definition  for connection  $G_{\alpha}$ in terms of generators $J^{ab}$.

\subsection*{9.  Bilinear combinations in a Riemannian space-time  }

This Section deals with methods for constructing  some bilinear invariant combinations in terms of
$\Phi$- and $\bar{\Phi}$-functions. We start with two equations
$$
[  \; \Gamma^{\alpha}(x)  \; ( \stackrel{\rightarrow}{\partial}_{\alpha} + \; G_{\alpha} - ig A_{\alpha} )
 - M \; ] \; \Phi = 0 \; ,
\eqno(9.1)
$$
$$
\bar{\Phi}  \;
[ \; ( \stackrel{\leftarrow}{\partial}_{\alpha} -\;
G_{\alpha} + ig A_{\alpha} ) \; \Gamma^{\alpha +} \; + \;
M \; ] = 0 \; .
\eqno(9.2)
$$

\noindent  Multiplying eq. (9.1) from the left by $\bar{\Phi}$, and eq.  (9.2) --- from the right by
$\Phi $, and  adding results together, we get to
$$
\bar{\Phi} \stackrel{\leftarrow}{\partial}_{\alpha}   \;
\Gamma^{\alpha} \Phi  \; +  \;
\bar{\Phi}  \Gamma^{\alpha} \; \stackrel{\rightarrow}{\partial}_{\alpha}
\Phi \; +
 \;\bar{\Phi}  \; ( \Gamma^{\alpha}  G_{\alpha}   \; -  \;
G_{\alpha} \Gamma^{\alpha}  ) \;  \Phi = 0 \; .
\eqno(9.3)
$$

To proceed with an analysis of eq. (9.3),   we need one auxiliary relation.To this end,  let us turn
again to a relativistic invariance condition for the Petras equation in matrix form\footnote{Its truth
can be easily proved with the use of spinor basis and properties  (3.4).}
$$
S \; \Gamma^{a} \; S^{-1} = \Gamma^{b} \;  L_{b}^{\;\;a} \; ,
\eqno(9.4)
$$

\noindent
and set a Lorentz transformation to be infinitesimal, from where it follows
$$
J^{kl}  \; \Gamma^{a} - \Gamma^{a} \; J^{kl} = \Gamma^{b} \;  (V^{kl})_{b}^{\;\;a} \; .
$$

\noindent Taking into account  the vector generators explicitly,
we arrive at a commutation relation
$$
J^{kl} \; \Gamma^{a} - \Gamma^{a} \; J^{kl} =
\Gamma^{k}\; g^{lk} - \Gamma ^{l} \; g^{ka} \; .
\eqno(9.5)
$$

\noindent
Now, multiplying eq. (9.5)  by an expression
$
e_{(a)}^{\rho} \; {1 \over 2} e_{(k)}^{\beta}
                  \nabla_{\sigma}  e_{(l)\beta } \; $ ,
we  obtain a needed formula
$$
\Gamma^{\rho}  \; G_{\sigma}  -
G_{\sigma}  \; \Gamma^{\rho}  =
\nabla_{\sigma} \;  \Gamma^{\rho}   \; .
\eqno(9.6)
$$

\noindent By accounting (9.6)  in (9.3),  we come to
$$
\bar{\Phi} \stackrel{\leftarrow}{\partial}_{\alpha}   \;
\Gamma^{\alpha}  \Phi  \; +  \;
\;\bar{\Phi}  \; ( \nabla_{\alpha} \Gamma^{\alpha} )  \;  \Phi +
\bar{\Phi}  \Gamma^{\alpha} \;
\stackrel{\rightarrow}{\partial}_{\alpha} \Phi  = 0 \; ,
\eqno(9.7a)
$$

\noindent
which can be rewritten as  a generally covariant conserved current law:
$$
\nabla_{\alpha} J^{\alpha} = 0 \; , \;
J^{\alpha} =   \; \bar{\Phi} \Gamma^{\alpha} \Phi    \; .
\eqno(9.7b)
$$

\noindent It is  easily verified that this conserved current $J_{\alpha}$ coincides with the above form (4.18).

It is particularly remarkable that the formula (9.7a) provides us with a hint  about  a general rule
for covariant $\nabla_{\alpha}$-differentiating any bilinear combination constructed on the base of
$\Phi$ and $\bar{\Phi}$ functions.

Indeed,  the rule
$$
\nabla_{\alpha}  \; ( \bar{\Phi} \Gamma^{\alpha}  \Phi ) =
\bar{\Phi} \; (\stackrel{\leftarrow}{\partial}_{\alpha} - G_{\alpha})  \;
\Gamma^{\alpha} \; \Phi  \; +  \;
\bar{\Phi}  \Gamma^{\alpha} \;
(\stackrel{\rightarrow}{\partial}_{\alpha}  +
  G_{\alpha}) \; \Psi  \;
\eqno(9.8a)
$$

\noindent
is in force. Further it will be convenient to use the notation
$$
(\stackrel{\rightarrow}{\nabla}_{\alpha}  + \; G_{\alpha} ) \Psi  =
\stackrel{\rightarrow}{D}_{\alpha}  \Phi \;  , \qquad
\bar{\Phi} (\stackrel{\leftarrow}{\nabla}_{\alpha}  - \; G_{\alpha} )   =
\bar{\Phi} \stackrel{\leftarrow}{D}_{\alpha} \; .
\eqno(9.8b)
$$

\noindent As usual, covariant derivatives  $\stackrel{\rightarrow}{\nabla}_{\alpha}$ and
$\stackrel{\leftarrow}{\nabla}_{\alpha} $ in acting on scalar functions become ordinary derivatives:
$\stackrel{\rightarrow}{\partial}_{\alpha}$ and
$\stackrel{\leftarrow}{\partial}_{\alpha} $ respectively.
With the notation (9.8b), the above eq. (9.8a) looks as
$$
\nabla_{\alpha} \; ( \bar{\Phi} \Gamma^{\alpha}  \Phi ) =
\bar{\Phi} \stackrel{\leftarrow}{D}_{\alpha} \; \Gamma^{\alpha} \Phi  \; +  \;
\bar{\Psi}  \Gamma^{\alpha} \;\stackrel{\rightarrow}{D}_{\alpha} \; \Phi  \; .
\eqno(9.8c)
$$

As prescribed by the formula (9.8c),  one  may replace the action of
$\nabla_{\alpha}$ by action of $\stackrel{\leftarrow}{D}_{\alpha}$  from  the right and  action of
$\stackrel{\rightarrow}{D}_{\alpha}$ from the left  respectively on $\bar{\Phi}$ and $\Phi$, completely
 ignoring a generally covariant nature of  the matrix $\Gamma^{\alpha}$ in the middle.

What is more, such a rule will work  always at any complicated bilinear function.  For instance,  let
us consider  a combination
$
\nabla^{\alpha} [ \; \bar{\Phi} \Gamma^{\rho}(x) \Gamma^{\sigma}(x) \Xi \; ]
\; $.
In accordance with  defining properties of the covariant derivative we can proceed
$$
\nabla^{\alpha} [ \; \bar{\Phi} \Gamma^{\rho} \Gamma^{\sigma} \Xi \; ] =
(  \partial_{\alpha} \bar{\Psi} ) \; \Gamma^{\rho} \Gamma^{\sigma} \; \Xi  +
\bar{\Phi} \; (\nabla_{\alpha} \Gamma^{\rho} ) \; \Gamma^{\sigma} \; \Xi +
 \bar{\Phi} \; \Gamma^{\rho} \; ( \nabla_{\alpha} \Gamma^{\sigma} ) \; \Xi  +
\bar{\Phi} \; \Gamma^{\rho}   \Gamma^{\sigma} \;
\; (\partial_{\alpha} \Xi ) \; .
$$

\noindent
Replacing the covariant derivatives of $\Gamma$-matrices with commutators as prescribed (9.6), one
gets
$$
\nabla^{\alpha} [ \; \bar{\Phi} \Gamma^{\rho} \Gamma^{\sigma} \Xi \; ] =
$$
$$
= ( \partial_{\alpha} \bar{\Phi})  \; \Gamma^{\rho} \Gamma^{\sigma} \; \Xi  +
\bar{\Phi} \; (\Gamma^{\rho} B_{\alpha}  - G_{\alpha} \Gamma^{\rho}  )\;
\Gamma^{\sigma} \; \Xi +
\bar{\Phi} \; \Gamma^{\rho} \; (\Gamma^{\sigma} G_{\alpha} - G_{\alpha} \Gamma^{\sigma}  )\;  \Xi  +
\bar{\Phi} \; \Gamma^{\rho} \Gamma^{\sigma}  \;( \partial_{\alpha} \Xi ) \; .
$$

\noindent Here two terms containing the product $\Gamma^{\rho} B_{\alpha}\Gamma^{\sigma}$ cancel each other,
 and  the remaining ones  can be rewritten as
$$
\nabla^{\alpha} \; ( \;  \bar{\Phi} \Gamma^{\rho} \Gamma^{\sigma} \; \Xi \; ) \;
 =  \bar{\Phi} \; \stackrel{\leftarrow}{D}_{\alpha} \;  \Gamma^{\rho}
 \Gamma^{\sigma}  \; \Xi + \;
\bar{\Phi} \;  \Gamma^{\rho} \Gamma^{\sigma}     \;
\stackrel{\rightarrow}{D}_{\alpha}  \; \Xi  \; .
\eqno(9.9a)
$$

\noindent In  turn, relation (9.9a) is equivalent to
$$
\nabla_{\alpha} \; (  \Gamma^{\rho} \Gamma^{\sigma} ) =
-\;  G_{\alpha}  \;  \Gamma^{\rho}   \Gamma^{\sigma} \;  + \;
\Gamma^{\rho} \Gamma^{\sigma}  \; G_{\alpha}
\; .
\eqno(9.9b)
$$

An analogous  formula for a bilinear  combination with any tensor structure follows immediately
by induction. Indeed, if
$$
\Gamma^{(n)} = \Gamma^{\rho_{1}}  \; \Gamma^{\rho_{2}}
... \;\Gamma^{\rho_{n}}  \; , \;\; \mbox{and} \;\;
\nabla_{\alpha} \; \Gamma^{(n)}  = \Gamma^{(n)} \; G_{\alpha} \; - \;
G_{\alpha} \; \Gamma^{(n)} \; ;
\eqno(9.10a)
$$

\noindent    then for $\Gamma^{(n+1)} = \Gamma^{(n)} \Gamma^{\rho} $  one  derives
$$
\nabla_{\alpha}   \; \Gamma^{(n+1)}  =
 - \; G_{\alpha} \;  \Gamma^{(n+1)} + \Gamma^{(n+1)} \; G_{\alpha}  \; .
\eqno(9.10b)
$$

\noindent From  (9.10) one can produce the following commutation
relations
$$
\Gamma^{(n) }  \stackrel{\rightarrow}{D_{\sigma}} =
         \stackrel{\rightarrow}{D_{\sigma}} \; \Gamma^{(n)} \; , \qquad
\Gamma^{(n)} \stackrel{\leftarrow}{D_{\sigma}} =
         \stackrel{\leftarrow}{D_{\sigma}}\; \Gamma^{(n)}\; .
\eqno(9.11)
$$

To close this Section, let us write down a Lagrange density for  the system under consideration. In matrix
approach that will be of the form
$$
L = + {1 \over 2} \; ( \bar{\Phi} \;\gamma^{\alpha} \; \stackrel{\rightarrow}{D}_{\alpha} \;
 \Phi  \; - \;
\bar{\Phi} \; \Gamma^{\alpha} \; \stackrel{\leftarrow}{D}_{\alpha} \; \Phi ) \; -\; m\;
 \bar{\Phi } \Phi \; .
\eqno(9.12)
$$

\subsection*{10. On canonical energy-momentum tensor}

In Sec. 10 we are going to  study a question about energy-momentum tensor for  a generalized
fermion. At this it is  convenient to exploit  a matrix formalism, when  two basic equations
 are given by (for much generality, an external  electromagnetic field will be  taken into account)
 $$
( \; \Gamma^{\alpha } \; \stackrel{\rightarrow}{D}_{\alpha}   \; - M \;  )
 \;\Phi   = 0 \; , \qquad
\bar{\Phi}  \;  (  \; \Gamma^{\alpha}  \stackrel{\leftarrow}{D}_{\alpha} \; +
M  \; ) = 0 \; ,
\eqno(10.1a)
$$

\noindent where
$$
\stackrel{\rightarrow}{D}_{\alpha} =  \nabla_{\alpha} + B_{\alpha}  -
i g A_{\alpha} \; , \qquad
\stackrel{\leftarrow}{D}_{\alpha} =
\stackrel{\leftarrow}{\nabla}_{\alpha} - B_{\alpha} + i
g  A_{\alpha} \; , \qquad  g \equiv e /\hbar c  \; .
\eqno(10.1b)
$$

Let us introduce  a tensor  quantity
$$
W^{\;\; \alpha}_{\beta} =  \bar{\Phi} \; \Gamma^{\alpha} \;
 \stackrel{\rightarrow}{D}_{\beta}  \; \Phi =
 \bar{\Phi} \; \Gamma^{\alpha} \;
( \stackrel{\rightarrow}{\partial}_{\beta} \; + \; B_{\beta}) \; \Phi  \;
- \; i g   A_{\beta} \; (\bar{\Psi} \; \Gamma^{\alpha} \;  \Phi  )\;
\eqno(10.2)
$$

\noindent and calculate a divergence over it. To this end,  acting on the first  equation in (10.1a) from the left
by  $\bar{\Psi} \stackrel{\rightarrow}{D}_{\beta}$, and simultaneously multiplying  the second equation in
(10.1a) from the right by  $\stackrel{\rightarrow}{D}_{\beta} \Phi$,  and summing the results, we get
$$
\bar{\Phi} \; \stackrel{\rightarrow}{D}_{\beta} \;
  \Gamma^{\alpha}  \stackrel{\rightarrow}{D}_{\alpha} \Phi  \; + \;
\bar{\Phi}  \stackrel{\leftarrow}{D}_{\alpha} \Gamma^{\alpha}  \;
\stackrel{\rightarrow}{D}_{\beta} \Phi  = 0  \; ,
\eqno(10.3a)
$$

\noindent and further
$$
\bar{\Phi} \;\Gamma^{\alpha} \; [ \; (\stackrel{\rightarrow}{D}_{\beta}
\stackrel{\rightarrow}{D}_{\alpha} - \stackrel{\rightarrow}{D}_{\alpha}
\stackrel{\rightarrow}{D}_{\beta}) \; + \;
\stackrel{\rightarrow}{D}_{\alpha} \stackrel{\rightarrow}{D}_{\beta}\;  ]\;
\Phi  \; +
\;\bar{\Phi} \stackrel{\leftarrow}{D}_{\alpha}  \Gamma^{\alpha} \;
\stackrel{\rightarrow}{D}_{\beta} \Phi  = 0 \; .
\eqno(10.3b)
$$

\noindent Now, placing a term with commutator
$[\stackrel{\rightarrow}{D}_{\beta}, \stackrel{\rightarrow}{D}_{\alpha}]_{-}$
on the right,  we get to
$$
\bar{\Phi} (\;   \Gamma^{\alpha} \stackrel{\rightarrow}{D}_{\alpha} \; + \;
\stackrel{\leftarrow}{D}_{\alpha}   \Gamma^{\alpha}\; ) \;
\stackrel{\rightarrow}{D}_{\beta}  \; \Phi   = \bar{\Phi} \; \Gamma^{\alpha} \;
[ \stackrel{\rightarrow}{D}_{\alpha} , \stackrel{\rightarrow}{D}_{\beta} ]_{-} \;
\Phi \; .
\eqno(10.3c)
$$

\noindent
Let us show that  an expression on the left  can be thought as  a  pure  divergence of
 $W^{\alpha}_{\;\;\beta}$.  We are to start with
$$
\bar{\Phi} (\;   \Gamma^{\alpha}
\stackrel{\rightarrow}{D}_{\alpha}  +
 \stackrel{\leftarrow}{D}_{\alpha}   \Gamma^{\alpha}\; ) \;
\stackrel{\rightarrow}{D}_{\beta}   \Phi   =
$$
$$
=
\bar{\Phi} [ \; \Gamma^{\alpha} (\stackrel{\rightarrow}{\nabla}_{\alpha} +
B_{\alpha} - i g  A_{\alpha})
 +
( \stackrel{\leftarrow}{\nabla}_{\alpha} - B_{\alpha} +
i g A_{\alpha})
\Gamma^{\alpha} \; ] \; \stackrel{\rightarrow}{D}_{\beta} \Phi =
$$
$$
= \bar{\Phi} \; [ \; \Gamma^{\alpha} \stackrel{\rightarrow}{\nabla}_{\alpha}
+  ( \Gamma^{\alpha}  B_{\alpha}
-  B_{\alpha} \Gamma^{\alpha}  )  +
\stackrel{\leftarrow}{\nabla}_{\alpha}
\Gamma^{\alpha} \; ] \;  \stackrel{\rightarrow}{D}_{\beta} \Phi   \; .
\eqno(10.4a)
$$

\noindent Taking into consideration  eq. (7.7),  relation (10.4a) can be transformed into
$$
\bar{\Phi} \; ( \;\Gamma^{\alpha}  \stackrel{\rightarrow}{D}_{\alpha}
+
\stackrel{\leftarrow}{D}_{\alpha}   \Gamma^{\alpha}
\; ) \; \stackrel{\rightarrow}{D}_{\beta}   \Phi   =
 \bar{\Phi} \; (\;
\stackrel{\leftarrow}{\nabla}_{\alpha}
\Gamma^{\alpha} \; +   \Gamma^{\alpha}_{\;\; ; \alpha} +
\Gamma^{\alpha}
  \stackrel{\rightarrow}{\nabla}_{\alpha} \; ) \;
\stackrel{\rightarrow}{D}_{\beta} \Phi  =
$$
$$
= \nabla_{\alpha}  ( \bar{\Phi}
\Gamma^{\alpha} \stackrel{\rightarrow}{D}_{\beta}
\Phi  )  \; .
\eqno(10.4b)
$$

\noindent
Hence,  eq.  (10.3b) is equivalent to
$$
\nabla_{\alpha}\; (\; W^{\;\;\;\alpha}_{\beta}\;  ) =
\bar{\Phi} \;  \Gamma^{\alpha}  \; [ \stackrel{\rightarrow}{D}_{\alpha} ,
\stackrel{\rightarrow}{D}_{\beta} ]_{-} \; \Phi \; .
\eqno(10.5)
$$

\noindent
Let us consider  in more detail the commutator
$$
[\stackrel{\rightarrow}{D}_{\alpha}, \stackrel{\rightarrow}{D}_{\beta} ]_{-} =
- i g  F_{\alpha \beta} \; + \;
D_{\alpha \beta} \; ,
\eqno(10.6a)
$$

\noindent where
$$
F_{\alpha \beta} = {\partial A_{\beta} \over \partial x^{\alpha }}  -
                   {\partial A_{\alpha}\over  \partial x^{\beta  }} \; , \qquad
D_{\alpha \beta} =   {\partial B_{\beta } \over \partial  x^{\alpha} }      -
                     {\partial B_{\alpha} \over \partial  x^{\beta } }
+  ( B_{\alpha} B_{\beta} - B_{\beta} B_{\alpha}  ) \; .
\eqno(10.6b)
$$

\noindent
For the first term in the expression for $D_{\alpha \beta}$, it is easily to obtain a representation
$$
\partial_{\alpha } B_{\beta}  -
\partial_{\beta }  B_{\alpha} = \nabla_{\alpha} B_{\beta} -
                                        \nabla_{\beta} B_{\alpha} =
$$
$$
= {1 \over 2} J^{ab} \nabla_{\alpha} \;
 ( \; e_{(a)}^{\nu} e_{(b)\nu ; \beta} \;) \; -   \;
  {1 \over 2} J^{ab} \nabla_{\beta}  \;
( \;e_{(a)}^{\nu} e_{(b)\nu ; \alpha} \; ) =
$$
$$
= {1 \over 2} J^{ab} e_{(a)}^{\nu} \; [\; e_{(b)\nu ; \beta ; \alpha } -
                                      e_{(b)\nu ; \alpha ; \beta }\; ] \; +
{1 \over 2 } J^{ab} \; [\; e_{(a) \nu ; \alpha } e_{(b) ; \beta}^{\nu} -
                      e_{(a) \nu ; \beta} e_{(b) ; \alpha}^{\nu} \; ] \; .
\eqno(10.7)
$$

\noindent
For second one it follows
$$
( B_{\alpha} B_{\beta} - B_{\beta} B_{\alpha}  ) =
$$
$$
=
(\; {1 \over 2} J^{ab} e_{(a)}^{\nu} e_{(b)\nu ; \alpha }  ) \;
(\; {1 \over 2} J^{kl} e_{(k)}^{\mu} e_{(l)\mu ; \beta } \; ) \; - \;
(\; {1 \over 2} J^{kl} e_{(k)}^{\mu} e_{(l)\mu ; \beta }  ) \;
(\; {1 \over 2} J^{ab} e_{(a)}^{\nu} e_{(b)\nu ; \alpha } \; ) \; =
$$
$$
= {1 \over 4} \left ( J^{ab} J^{kl}    -  J^{kl}  J^{ab} \right  ) \left [
\;   (  e_{(a)}^{\nu} e_{(b)\nu ; \alpha }  ) \;
\; e_{(k)}^{\mu} e_{(l)\mu ; \beta } \; )  \right ] \; .
$$

\noindent Now,  with the use of the  commutation relation (see (6.11))
$$
[J^{ab},J^{kl}]_{-} =
( -J^{kb} g^{la} + J^{lb} g^{ka} ) - (  - J^{ka} g^{lb}  + J^{la} g^{kb} ) \; ,
$$

\noindent we get
$$
( B_{\alpha} B_{\beta} - B_{\beta} B_{\alpha}  ) =
 - {1 \over 2 } J^{ab} \; [\; e_{(a) \nu ; \alpha } e_{(b) ; \beta}^{\nu} -
                      e_{(a) \nu ; \beta} e_{(b); \alpha}^{\nu} \; ] \; .
\eqno(10.8)
$$

\noindent By accounting eqs. (10.7) and (10.8),  the above expression for $D_{\alpha \beta}$
is led to
$$
D_{\alpha \beta} =  {1 \over 2} J^{ab}\; e_{(a)}^{\nu} \;
[\; e_{(b)\nu ; \beta ; \alpha } - e_{(b)\nu ; \alpha ; \beta }\; ] =
$$
$$
=
{1 \over 2} J^{ab}\; e_{(a)}^{\nu} \; [ \;e_{(b)} ^{\rho} R_{\rho \nu \beta \alpha} (x)\; ] =
{1 \over 2 } J^{\nu \rho} (x)\; R_{\nu \rho \alpha \beta }(x) \; ,
\eqno(10.9)
$$

\noindent where $R_{\nu \rho \alpha \beta }(x)$  is the Riemann curvature tensor.
So, eq.  (10.5) reads as
$$
\nabla_{\alpha} W^{\;\;\;\alpha}_{\beta}  =
- ig \; J^{\alpha}    F_{\alpha \beta }               \; + \;
{1 \over 2} \; R_{\nu \rho \alpha \beta}  \;
\bar{\Phi} \Gamma^{\alpha}  J^{\nu \rho} \Phi  \; .
\eqno(10.10)
$$

\noindent
In is useful to perform  some transformation over second term on the right:
$$
{1 \over 2} \; R_{\nu \rho \alpha \beta}  \;
\bar{\Phi} \; \Gamma^{\alpha} J^{\nu \rho} \; \Phi  =
$$
$$
=
{1 \over 2} \; R_{\nu \rho \alpha \beta}  \; \bar{\Phi} \;
\left ( {1 \over 2} [\;
\Gamma^{\alpha} J^{\nu \rho} -
J^{\nu \rho}  \Gamma^{\alpha}\;  ]  +
{1 \over 2}  [\; \Gamma^{\alpha} J^{\nu \rho} +
J^{\nu \rho} \Gamma^{\alpha}  \; ] \right ) \; \Phi \; .
$$

\noindent Further,  with the use of commutation relation (see (7.6))
$$
\Gamma^{\alpha} J^{\nu \rho} -
J^{\nu \rho}  \Gamma^{\alpha} =
g^{\alpha \nu}(x) \Gamma^{\rho} - g^{\alpha \rho} (x) \Gamma^{\nu} \; ,
$$

\noindent one can  produce
$$
{1 \over 2} \; R_{\nu \rho \alpha \beta}  \;
\bar{\Phi} \; \Gamma^{\alpha} J^{\nu \rho} \; \Phi  =
$$
$$
=
{1 \over2 } R_{\alpha \beta} J^{\alpha}  + {1 \over 4}
\; R_{\nu \rho \alpha \beta}  \; \bar{\Phi} \;
(\Gamma^{\alpha}  J^{\nu \rho} +
J^{\nu \rho} \Gamma^{\alpha}  ) \; \Phi \; .
\eqno(10.11)
$$

\noindent Correspondingly, eq.  (10.10) takes on the form
 $$
\nabla_{\alpha} \; [ \; W^{\;\;\;\alpha}_{\beta} \; ]  =
J^{\alpha}  [ \;-i g \; F_{\alpha \beta }\; + \;
{1 \over2 }\;  R_{\alpha \beta} \; ]\; + \;  {1 \over 4}
\; R_{\nu \rho \alpha \beta}  \; \bar{\Phi} \;
(\Gamma^{\alpha}  J^{\nu \rho} + J^{\nu \rho} \Gamma^{\alpha}  )\; \Phi \; ,
\eqno(10.12)
$$

\noindent which generalizes a formula established by V.A. Fock  ( [49], eq. (56)) at studying
a spin 1/2 particle on the background of a curved space-time model.
A single formal difference  consists  in occurrence  of one  additional  term proportional
to the curvature tensor.

It may be checked that, in  the established formula (10.12) being applied to  an ordinary spin 1/2 particle,
 this additional $R$-dependent term will vanish identically. To this end,  let us  consider  more closely
 a combination of the Dirac matrices
 $$
(\gamma^{\alpha}  \sigma^{\nu \rho} + \sigma^{\nu \rho} \gamma^{\alpha}  )
= {1 \over 4}
[\; \gamma^{\alpha} (\gamma^{\nu} \gamma^{\rho} - \gamma^{\rho} \gamma^{\nu}) +
(\gamma^{\nu} \gamma^{\rho} - \gamma^{\rho} \gamma^{\nu})
\gamma^{\alpha}  \; ] \; .
$$

\noindent We need one auxiliary  relation. To produce it, one should multiply the known formula for
the product of  three Dirac matrices [12]
$$
\gamma^{a} \gamma^{b} \gamma^{c} =
\gamma^{a} g^{bc} -  \gamma^{b} g^{ac} + \gamma^{c} g^{ab} +
i \gamma^{5} \epsilon^{abcd} \gamma_{d}
\eqno(10.13a)
$$

\noindent  by  a tetrad-based expression
$e^{\alpha}_{(a)} e^{\beta}_{(b)} e^{\rho}_{(c)}$, so that we get to
$$
\gamma ^{\alpha }(x) \; \gamma ^{\beta }(x) \; \gamma ^{\rho }(x)  =
[\;  \gamma ^{\alpha }(x) \; g^{\beta \rho }(x) \;  -  \;
\gamma ^{\beta }(x) \; g^{\alpha \rho }(x) \; +
$$
$$
+ \; \gamma ^{\rho }(x) \; g^{\alpha \beta }(x) \; + \; i \gamma ^{5} \;
\epsilon ^{\alpha \beta \rho \sigma }(x)\; \gamma _{\sigma }(x)
\; ] \;  .
\eqno(10.13b)
$$

\noindent Here, a generally covariant Levi-Civita symbol is defined by
$$
\epsilon ^{\alpha \beta \rho \sigma }(x) =
e^{\alpha}_{(a)}   e^{\beta}_{(b)}  e^{\rho}_{(c)}   e^{\sigma}_{(d)}
\epsilon^{abcd}  \; .
\eqno(10.13c)
$$

\noindent Now, with the use of eq. (10.13b), we readily produce
$$
\gamma^{\alpha} (x) \sigma^{\nu \rho} (x) + \sigma^{\nu \rho}(x)
\gamma^{\alpha} (x) =
i \gamma^{5}\; \epsilon^{\alpha \nu \rho \sigma}(x) \; \gamma_{\sigma}(x) \; .
\eqno(10.14a)
$$

\noindent Thus, the following relation
$$
{1 \over 4}
\; R_{\nu \rho \alpha \beta} (x) \; \bar{\Phi}
(\Gamma^{\alpha}  \sigma^{\nu \rho} +
\sigma^{\nu \rho} \Gamma^{\alpha}  )\Phi =
{1 \over  4}  \; R_{\nu \rho \alpha \beta}(x)  \;
 \epsilon^{\alpha \nu \rho \sigma}(x) \;
\bar{\Phi}    \gamma^{5}  \; \gamma_{\sigma}(x)  \Phi  \equiv 0 \;
\eqno(10.14b)
$$

\noindent holds; at deriving eq. (10.14b) we have taken into account the known symmetry
property of the  curvature tensor under  cyclic permutation over any three indices so that a
convolution $R^{....}$ with  $\epsilon_{....}$ over three indices equals to zero.

Returning again to eq. (10.12a), and bearing in mind the properties of the $H$-matrix
$$
 H^{-1} [\Gamma^{\alpha}(x)]^{+}  H = - \; \Gamma^{\alpha}(x) \; , \qquad
H^{-1} [J^{\nu \rho}(x)]^{+} H = - J^{\nu \rho }(x)  \; ,
\eqno(10.15)
$$

\noindent one straightforwardly finds that  on the right in (10.12a) the first and third terms are real-valued,
whereas the second one is imaginary:
$$
\nabla_{\alpha} W^{\;\;\;\alpha}_{\beta}  =
Re (x) +  i \; Im (x) \;, \;\;
Im (x) =
-   \;
{i \over2 }   J^{\alpha}  R_{\alpha \beta}  \;  ,
$$
$$
Re (x) =
 \;- i g \; J^{\alpha}   F_{\alpha \beta }      \; + \; {1 \over 4 }
\; R_{\nu \rho \alpha \beta}  \; \bar{\Phi}
(\Gamma^{\alpha}  J^{\nu \rho} +
J^{\nu \rho} \beta^{\alpha}  ) \Phi \; .
\eqno(10.16)
$$

\noindent
Now,  we need to isolate  real and  imaginary parts on the left in (10.16) too. With  this aim in mind,
one is to find a complex conjugate tensor  $(W^{\alpha} _{\;\;\beta})^{*}$:
$$
(W^{\;\;\;\alpha}_{\beta})^{+} = [ \; \Phi^{+} \eta \Gamma^{\alpha}
\; (\nabla_{\beta} + G_{\beta} - ig A_{\beta}) \; \Phi \; ]^{+} =
$$
$$
=
- \; \bar{\Phi}  \; \Gamma^{\alpha} \;
(\stackrel{\leftarrow}{\nabla}_{\beta} - G_{\beta} + ig A_{\beta}) \; \Phi =
- \; \bar{\Phi} \stackrel{\leftarrow}{D}_{\beta} \Gamma^{\alpha} \;  \Phi \; .
\eqno(10.17)
$$

\noindent With the notation
$$
Re \;( W^{\;\;\; \alpha}_{\beta} ) = {1 \over 2} \left
  [ \; W^{\;\;\;\alpha}_{\beta} +
    ( W^{\;\;\; \alpha}_{\beta} )^{+} \; \right ] = T^{\;\;\;\alpha}_{\beta} \; ,
$$
$$
Im  \; ( W^{\;\;\; \alpha}_{\beta} ) = {1 \over 2 i } \left
   [ \; W^{\;\;\; \alpha}_{\beta} -
     (  W^{\;\;\; \alpha}_{\beta} )^{+} \; \right ] =
U^{\;\;\; \alpha}_{\beta}\; .
\eqno(10.18)
$$

\noindent eq.  (10.16) will split into two real-valued ones:
$$
\nabla_{\alpha} (T^{\;\;\; \alpha}_{\beta}) =
- i g \; J^{\alpha}(x)  F_{\alpha \beta}    \; + \; {1 \over 4 }
\; R_{\nu \rho \alpha \beta}  \; \bar{\Phi} \;
(\Gamma^{\alpha}  J^{\nu \rho} +
J^{\nu \rho} \Gamma^{\alpha}  ) \; \Phi  \; ,
\eqno(10.19)
$$
$$
\nabla_{\alpha} (U^{\;\;\; \alpha}_{\beta}) =
-{i \over2 }  J^{\alpha } R_{\alpha \beta}  \; .
\eqno(10.20)
$$

As readily checked, eq. (10.20) represents in essence a direct consequence of the conserved current law.
Indeed,  in accordance with definition for  $U^{\;\;\; \alpha}_{\beta}$,  we have
$$
U^{\;\;\; \alpha}_{\beta} = {1 \over 2i} \left [
\bar{\Phi}
\Gamma^{\alpha} ( \partial_{\beta} + G_{\beta} - i g A_{\beta} ) \Phi +
\bar{\Phi} ( \stackrel{\leftarrow}{\partial}_{\beta} -
G_{\beta} + i g A_{\beta}) \Gamma^{\alpha} \Phi
\right ] =
$$
$$
= {1 \over 2 i } \nabla_{\beta} \; ( \bar{\Psi} \Gamma^{\alpha} \Psi ) =
{1 \over 2i} \nabla_{\beta} J^{\alpha}\; .
\eqno(10.21)
$$

\noindent
Therefore, eq. (10.19)  can be led to the form
$$
\nabla_{\alpha} \nabla_{\beta} J^{\alpha} =
R_{\alpha \beta} J^{\alpha}  \; ,
\eqno(10.22a)
$$

\noindent and further
$$
(\nabla_{\alpha} \nabla_{\beta}    - \nabla_{\beta} \nabla_{\alpha}) J^{\alpha} \; + \;
\nabla_{\beta} \nabla_{\alpha} J^{\alpha}  =
 J^{\alpha} \; R_{\alpha \beta}   \; .
$$

\noindent From this,  with  bearing in mind  the  conservation law for  $J^{\alpha}$,  it follows immediattely
an identity
$$
J^{\rho} (x) \;  R_{\rho\;\;\;    \beta \alpha }^{\;\;\alpha}(x)  \equiv
 J^{\alpha}(x)  \;  R_{\alpha \beta} (x)   \; .
\eqno(10.22б)
$$

\noindent Thus,  eq. (10.20)  does not include  anything new  in addition to  current conservation
law. As for eq.  (10.19),  we have
$$
T^{\;\;\;\alpha}_{\beta} =
{1 \over 2} [ \bar{\Phi} \Gamma^{\alpha} \stackrel{\rightarrow}{D}_{\beta} \Phi -
\bar{\Phi} \Gamma^{\alpha} \stackrel{\leftarrow}{D}_{\beta} \Phi ] =
$$
$$
= {1 \over 2 } [ \bar{\Phi} \Gamma^{\alpha}
(\stackrel{\rightarrow}{\nabla}_{\beta} + G_{\beta}) \Phi -
\bar{\Phi} \Gamma^{\alpha} (\stackrel{\leftarrow}{\nabla}_{\beta} -
 G_{\beta}) \Phi ]  -   ig J^{\alpha} A_{\beta} \; ,
\eqno(10.23a)
$$

\noindent  and  a conservation law  reads as follows
$$
\nabla_{\alpha} (T^{\;\;\; \alpha}_{\beta}) =
- i g \; J^{\alpha}  F_{\alpha \beta}    \; + \;
{1 \over 4 } \; R_{\nu \rho \sigma \beta}  \; \bar{\Phi} \;
(\Gamma^{\sigma}  J^{\nu \rho} + J^{\nu \rho} \Gamma^{\sigma}  ) \; \Phi  \; .
\eqno(10.23b)
$$

To proceed  with eq. (10.23b), it is a time to have remembered one property of the energy-momentum
tensor in Minkowski space-time, which  concerns its ambiguity in determination.
In the Minkowski space-time,  such a freedom is described as follows:
if $ T_{b}^{\;\;a}(x)$ obeys the conservation law
$$
 \partial_{a} T_{b}^{\;\;a} = 0 \; ,
$$

\noindent
then  another one
$$
\bar{T}_{b}^{\;\;a}(x) = T_{b}^{\;\;a}(x) + \partial_{c} \; [ \;
 \Omega_{b}^{\;\;[ac]}(x) \; ] \; , \qquad
\Omega_{b}^{\;\;[ac]}(x) = - \; \Omega_{b}^{\;\;[ca]}(x) \;
\eqno(10.24a)
$$

\noindent satisfies the same conservation law as well
$$
 \partial_{a} \bar{T}_{b}^{\;\;a} = 0 \; .
$$

\noindent Obviously,  the simultaneous existence of the two tensors  is insured by an elementary
formula
$$
\partial_{a} \partial_{c} \;  \Omega_{b}^{\;\;[ca]} (x) \equiv  0 \; .
\eqno(10.24b)
$$

In the case of a curved space-time such an equivalence between tensors  $T_{\beta}^{\;\; \alpha}(x) $ and
$\bar{T}_{\beta}^{\;\; \alpha }(x)$ holds as well, but in a more complicated manner. Indeed,
let two tensors be  related to each other by
$$
\bar{T}_{\beta}^{\;\; \alpha } (x) = T_{\beta}^{\;\; \alpha }(x) +
\nabla_{\rho} \; [ \; \Omega_{\beta}^{\;\; [\alpha \rho]} (x) \; ] \; .
\eqno(10.25)
$$

\noindent Acting on  both sides by operation of covariant derivative  $\nabla_{\alpha}$,  one produces
$$
\nabla_{\alpha}  \bar{T}_{\beta}^{\;\; \alpha } (x) =
\nabla_{\alpha}  T_{\beta}^{\;\; \alpha } (x)  \; +
\; \nabla_{\alpha } \; [ \; \nabla_{\rho} \Omega_{\beta}^{\;\; [\alpha \rho]} (x) \; ]\; .
\eqno(10.26)
$$

\noindent
Now,  bearing in mind symmetry properties of  the curvature tensor, one obtains
$$
\nabla_{\alpha} \; [ \;\nabla_{\rho} \Omega_{\beta}^{\;\; [\alpha \rho]} (x)\; ] =
{1 \over 2}\; [ \; R_{\alpha \rho \beta}^{\;\;\;\;\;\;\sigma}
 \; \Omega_{\sigma}^{\;\;[\alpha \rho]} +
R_{\alpha \rho\;\;\;\sigma}^{\;\;\;\;\alpha} \; \Omega_{\beta}^{\;\;[\sigma \rho]} +
R_{\alpha \rho\;\;\;\sigma}^{\;\;\;\;\rho} \; \Omega_{\beta}^{\;\;[\alpha \sigma]} \; ] =
$$
$$
= {1 \over 2} \; [\; R^{\beta \sigma}_{\;\;\;\;\; \alpha \rho} \;
\Omega_{\sigma}^{\;\;[\; \alpha \rho]} -
 R_{\rho \sigma} \; \Omega_{\beta}^{\;\;[\sigma \rho]} -
R_{\alpha \sigma} \;  \Omega_{\beta}^{\;\;  [\alpha \sigma]} \; ] \; ,
$$

\noindent therefore
$$
\nabla_{\alpha} \; [ \;\nabla_{\rho} \;\Omega_{\beta}^{\;\; [\alpha \rho]} (x)\; ] =
{1 \over 2} \; R_{\beta\sigma \nu \rho} \; \Omega^{\sigma[\nu \rho]} \; .
\eqno(10.27)
$$

\noindent Thus, eq.  (10.26) reads as
$$
\nabla_{\alpha} \; \bar{T}_{\beta}^{\;\;\; \alpha} (x) =
\nabla_{\alpha}  \; T_{\beta}^{\;\;\;  \alpha } (x)  +
{1 \over 2} \; R_{\beta\sigma \nu \rho} (x)
\; \Omega^{\sigma[\nu \rho]}(x) \; ,
\eqno(10.28)
$$

\noindent which on accounting eq. (10.23b) takes the form
$$
\nabla_{\alpha} \;  \bar{T}_{\beta}^{\;\; \alpha } (x) =
 -i g J^{\alpha} \; F_{\alpha \beta} \;  + \;
 {1 \over 4} \; R_{\nu \rho  \sigma \beta } (x) \;
\bar{\Phi} \; (\Gamma^{\sigma} J^{\nu \rho} + J^{\nu \rho} \Gamma^{\sigma } ) \;
\Phi \; + \;
$$
$$
+ \;
{1 \over 2}  \; R_{\beta \sigma \nu \rho} (x) \; \Omega^{\sigma [ \nu \rho]} (x) \; .
\eqno(10.29a)
$$

\noindent  If the quantity $\Omega^{\sigma [\nu \rho]}(x)$ is hcosen as
$$
\Omega^{\sigma [\nu \rho]}(x) =
+ {1 \over 2} \; \bar{\Phi} \;  (\; \Gamma^{\sigma} J^{\nu \rho} + J^{\nu \rho}
\Gamma^{\sigma} \; ) \;  \Phi \; ,
$$

\noindent
then second and  third terms on the right  in (10.29a) cancel each other, and  we reach a conservation
law in the form
$$
\nabla_{\alpha} \; \bar{T}_{\beta}^{\;\; \alpha } (x) = -i g \; J^{\alpha} (x) \;
F_{\alpha \beta } (x)  \; .
\eqno(10.29b)
$$

\subsection*{ Supplement A.
Ricci tensor and  conformal transformation}

The Ricci tensor is expressed in terms of Kristofell symbols
$$
R_{\alpha \beta} =
\partial_{\rho} \Gamma^{\rho}_{\alpha \beta} - \partial_{\beta}
\Gamma^{\rho}_{\alpha \rho} +
\Gamma^{\rho}_{\alpha \beta} \Gamma^{\sigma}_{\rho \sigma}
- \Gamma^{\sigma}_{\alpha \rho} \Gamma^{\rho}_{\sigma \beta} \; .
\eqno(A.1)
$$

\noindent
Let metric tensors of two space-time models differ by a factor
$$
dS^{2} = g_{\alpha \beta} dx^{\alpha} dx^{\beta} \; , \;
d\tilde{S}^{2} = \tilde{g}_{\alpha \beta} dx^{\alpha} dx^{\beta} \; , \;
$$
$$
\tilde{g}_{\alpha \beta}(x) = \varphi^{2} (x) \; g_{\alpha \beta} (x) \; .
\eqno(A.2)
$$

\noindent
Comparing  two set Kristoffel symbols
$$
\tilde{\Gamma}_{\alpha \beta, \rho} (x) =
{1 \over 2} \; [\;
\partial_{\alpha} \tilde{g}_{\beta \rho} + \partial_{\beta} \tilde{g}_{\alpha \rho}
- \partial_{\rho} \tilde{g}_{\alpha \beta} \; ] =
{1 \over 2} \; [ \;
\partial_{\alpha} \varphi^{2} g_{\beta \rho} +
\partial_{\beta} \varphi^{2} g_{\alpha \rho}
- \partial_{\rho} \varphi^{2} g_{\alpha \beta}  ) =
$$
$$
= \varphi^{2} \Gamma_{\alpha \beta, \rho} + \varphi \; [
(\partial_{\alpha} \varphi ) g_{\beta \rho} +
(\partial_{\beta} \varphi ) g_{\alpha \rho}
- (\partial_{\rho} \varphi ) g_{\alpha \beta} \; ] \; ,
$$

\noindent or
$$
\tilde{\Gamma}^{\sigma}_{\alpha \beta}(x) =
\tilde{g}^{\sigma \rho}
\Gamma_{\alpha \beta, \rho} = { 1 \over \varphi^{2}} g^{\sigma \rho}
\{ \varphi^{2} \Gamma_{\alpha \beta, \rho} + \varphi \;
[(\partial_{\alpha} \varphi ) g_{\beta \rho} +
(\partial_{\beta} \varphi ) g_{\alpha \rho}
- (\partial_{\rho} \varphi ) g_{\alpha \beta} \; ] \} \; .
$$

\noindent Thus we arrive at
$$
\tilde{\Gamma}^{\sigma}_{\alpha \beta} =
\Gamma^{\sigma}_{\alpha \beta} + { 1 \over \varphi}
[(\partial_{\alpha} \varphi ) \delta^{\sigma}_{\beta}  +
(\partial_{\beta} \varphi ) \delta^{\sigma}_{\alpha}
- g^{\sigma \rho }  (\partial_{\rho} \varphi ) g_{\alpha \beta} \; ] \} \; .
\eqno(A.3)
$$

Now let us compare the Ricci tensors
$$
\partial_{\rho} \tilde{\Gamma}^{\rho}_{\alpha \beta} =
\partial_{\rho} \Gamma^{\rho}_{\alpha \beta} -
$$
$$
- {1 \over \varphi^{2}} (\partial_{\rho} \varphi)
[\;  \delta^{\rho}_{\alpha} \partial_{\beta} \varphi   +
\delta^{\rho}_{\beta} \partial_{\alpha} \varphi  -
g_{\alpha \beta}(x) g^{\rho \sigma} \partial_{\sigma} \varphi \; ] +
$$
$$
+  {1 \over \varphi}
[\; 2 \partial_{\alpha} \partial_{\beta} \varphi - g_{\alpha \beta} g^{\rho \sigma}
\partial_{\rho} \partial_{\sigma} \varphi -
(\partial_{\rho} g_{\alpha \beta}) g^{\rho \sigma} \partial_{\sigma} \varphi -
g_{\alpha \beta} (\partial_{\rho} g^{\rho \sigma} ) \partial_{\sigma} \varphi \;]
$$

\noindent
and further
$$
\partial_{\rho} \tilde{\Gamma}^{\rho}_{\alpha \beta} =
\partial_{\rho} \Gamma^{\rho}_{\alpha \beta} -
$$
$$
- {1 \over \varphi^{2}} (\partial_{\rho} \varphi)
[ \; 2 \partial_{\alpha} \varphi  \partial_{\beta} \varphi
- g_{\alpha \beta}(x) g^{\rho \sigma}(x) \partial_{\rho} \varphi
\partial_{\sigma} \varphi \; ] +
$$
$$
+  {1 \over \varphi}
[\; 2 \partial_{\alpha} \partial_{\beta} \varphi -
g_{\alpha \beta} g^{\rho \sigma} \partial_{\rho} \partial_{\sigma} \varphi \; ]
-
$$
$$
-  {1 \over \varphi}
[ \; (\partial_{\rho} g_{\alpha \beta}) g^{\rho \sigma}  +
g_{\alpha \beta} (\partial_{\rho} g^{\rho \sigma} \; ]
 \partial_{\sigma} \varphi \; .
\eqno(A.4)
$$

Taking into account
$$
-\partial_{\beta} \tilde{\Gamma}^{\rho}_{\alpha \rho} =
- \partial_{\beta} ( \Gamma^{\rho}_{\alpha \rho} + { 4 \over \varphi}
\partial_{\alpha} \varphi ) =
$$
$$
= - \partial_{\beta}  \Gamma^{\rho}_{\alpha \rho}  +
{4 \over \varphi^{2}} (\partial_{\alpha} \varphi ) (\partial_{\beta} \varphi) -
{4 \over \varphi} \partial_{\alpha}  \partial_{\beta} \varphi \; .
\eqno(A.5)
$$

\noindent we  get
$$
\tilde{\Gamma}^{\rho}_{\alpha \beta} \tilde{\Gamma}^{\sigma}_{\rho \sigma} =
[ \; \Gamma^{\rho}_{\alpha \beta}  + {1 \over \varphi }
(\delta_{\alpha}^{\rho} \partial_{\beta} \varphi +
\delta^{\rho}_{\beta} \partial_{\alpha} \varphi -
g_{\alpha \beta} g^{\rho \gamma} \partial_{\gamma} \varphi ) \; ]\;
( \Gamma^{\sigma}_{\rho \sigma} + {4 \over \varphi }
\partial_{\rho} \varphi ) =
$$
$$
= \Gamma^{\rho}_{\alpha \beta} \Gamma^{\sigma} _{\rho \sigma}  +
{1 \over \varphi} [ \; 4 (\partial_{\rho} \varphi) \Gamma^{\rho}_{\alpha \beta} +
(\partial_{\beta} \varphi ) \Gamma^{\sigma} _{\alpha \sigma}  +
(\partial_{\alpha} \varphi ) \Gamma^{\sigma} _{\beta \sigma} -
g_{\alpha \beta} \Gamma^{\sigma}_{\rho \sigma} g^{\rho \gamma}
 (\partial_{\gamma}\varphi) \; ] +
$$
$$
+ {4 \over \varphi^{2}} [ \; 2 (\partial_{\alpha} \varphi)
(\partial_{\beta} \varphi) - g_{\alpha \beta} \; g^{\rho \sigma}
(\partial_{\rho} \varphi ) ( \partial_{\sigma} \varphi )  \; ] \; .
\eqno(A.6)
$$

Now
$$
- \tilde{\Gamma}^{\sigma}_{\alpha \rho}
\tilde{\Gamma}^{\rho}_{\sigma \beta} =
- \Gamma^{\sigma}_{\alpha \rho}
\Gamma^{\rho}_{\sigma \beta} -
$$
$$
- {1 \over \varphi^{2}}
[ \; 6 (\partial_{\alpha} \varphi) (\partial_{\beta} \varphi )
- 2 g_{\alpha \beta} g^{\rho \sigma} (\partial_{\rho} \varphi)
                                     (\partial_{\sigma} \varphi) -
$$
$$
-
{1 \over \varphi }
[ \; (\partial_{\beta} \varphi) \gamma^{\sigma} _{\alpha \sigma} +
(\partial_{\alpha} \varphi)  \Gamma^{\sigma}_{\beta\sigma} +
$$
$$
+ 2 (\partial_{\rho} \varphi ) \Gamma^{\rho}_{\alpha \beta} -
(\partial_{\gamma} \varphi)
g^{\Gamma \rho}   \Gamma^{\sigma}_{\rho \alpha}  g_{\sigma\beta}-
(\partial_{\gamma} \varphi)  g^{\gamma \rho} \Gamma^{\sigma}_{\rho \beta}
g_{\sigma\alpha} \; ]
\eqno(A.7)
$$

Allowing for eqs. (A.4)-(A.7), we get to
$$
\tilde{R}_{\alpha \beta} = R_{\alpha \beta} +
{1 \over \varphi } [ \;- 2 \partial_{\alpha} \partial_{\beta} \varphi -
g_{\alpha \beta} g^{\rho \sigma} \partial_{\rho} \partial_{\sigma} \varphi \; ]
+
$$
$$
+
{1 \over \varphi^{2}}
[ \; 4  (\partial_{\alpha} \varphi) ( \partial_{\beta} \varphi) -
g_{\alpha \beta} g^{\rho \sigma} (\partial_{\rho} \varphi )
(\partial_{\sigma} \varphi )\; ]+
$$
$$
+  {1 \over \varphi } [ \;
- (\partial_{\rho} g_{\alpha \beta})  g^{\rho \sigma} \partial_{\sigma} \varphi -
g_{\alpha \beta} (\partial_{\rho} g^{\rho \sigma}) \partial_{\sigma} \varphi +
2 (\partial_{\rho} \varphi ) \Gamma^{\rho}_{\alpha \beta} -
$$
$$
- g_{\alpha \beta} \Gamma^{\sigma}_{\rho \sigma}g^{\rho \gamma}
\partial_{\gamma} \varphi +
(\partial_{\gamma} \varphi ) g^{\gamma \rho} \Gamma^{\sigma}_{\rho \alpha}
g_{\sigma \beta} + (\partial_{\gamma} \varphi )g^{\gamma \rho}
\Gamma^{\sigma} _{\rho \beta} g_{\sigma \alpha} \; \
\eqno(A.8)
$$

Transforming here first two terms
$$
- (\partial_{\sigma} \varphi )   g^{\rho \sigma}  (\partial_{\rho} g_{\alpha \beta})  =
(\partial_{\sigma} \varphi)  g^{\sigma \rho} \; [
\Gamma_{\alpha,\beta \rho} + \Gamma_{\beta, \alpha \rho} ] =
$$
$$
=
(\partial_{\sigma} \varphi)  g^{\sigma \rho} \;
[\; g_{\alpha \gamma} \Gamma^{\gamma}_{\beta \rho} +
g_{\beta \gamma} \Gamma^{\gamma}_{\alpha \rho} \; ] =
$$
$$
= - ( \partial_{\sigma} \varphi ) g^{\sigma \rho} \Gamma^{\gamma}_{\rho \beta }
g_{\gamma \alpha} -
( \partial_{\sigma} \varphi ) g^{\sigma \rho} \Gamma^{\gamma}_{\rho \alpha }
g_{\gamma \beta}  \; ,
$$
$$
- g_{\alpha \beta} (\partial _{\sigma} \varphi )\; (\partial_{\rho} g^{\rho\sigma}) =
- g_{\alpha \beta} (\partial _{\sigma} \varphi )\;
[ \; - \Gamma^{\rho}_{\rho \gamma} g^{\gamma \sigma} -
\Gamma^{\sigma}_{\rho \gamma}  g^{\rho \gamma} \; ] =
$$
$$
=
+ g_{\alpha \beta} (\partial _{\sigma} \varphi )\;
\Gamma^{\rho}_{\rho \gamma} g^{\gamma \sigma}
+ g_{\alpha \beta} (\partial _{\sigma} \varphi )\;
\Gamma^{\sigma}_{\rho \gamma}  g^{\rho \gamma} \; .
$$

\noindent whith which eq. (A.8) gives
$$
\tilde{R}_{\alpha \beta} = R_{\alpha \beta} +
{1 \over \varphi } \; ( \;- 2 \partial_{\alpha} \partial_{\beta} \varphi -
g_{\alpha \beta} g^{\rho \sigma} \partial_{\rho} \partial_{\sigma} \varphi \; )
+
$$
$$
+
{1 \over \varphi^{2}}
[ \; 4  (\partial_{\alpha} \varphi) ( \partial_{\beta} \varphi) -
g_{\alpha \beta} g^{\rho \sigma} (\partial_{\rho} \varphi )
(\partial_{\sigma} \varphi )\; ]+
$$
$$
+  {1 \over \varphi } (\partial_{\sigma} \varphi) \; ( \;2
\Gamma^{\sigma} _{\alpha \beta} + g_{\alpha \beta} \Gamma^{\sigma}_{\gamma \rho} g^{\gamma \rho} \; ) \; .
\eqno(A.9)
$$

From eq. (A.9) it follows the transformation law:
$$
\tilde{R} = {1 \over \varphi^{2}} \{
R - {6 \over \varphi } \; g^{\alpha \beta} \; [ \;
\partial_{\alpha}  (\partial_{\beta} \varphi ) -
\Gamma^{\sigma} _{\alpha \beta}
(\partial_{\sigma} \varphi) ] \}=
{1 \over \varphi^{2}}  (
R -  {6 \over \varphi }\; \nabla^{\beta} \nabla_{\beta} \varphi  )
\eqno(A.10)
$$

\newpage

\begin{center}
{\bf References}
\end{center}

\noindent 1.
Pauli W., Fierz M. {\em Uber relativistische
Feldgleichungen von Teilchen mit beliebigem Spin im
electro\-magnetishen Feld.}  Helv. Phys. Acta. 1939, Bd. 12, S.
297-300.

\noindent
2.  Fierz V., Pauli W. {\em On relativistic wave equations
for particles of arbitrary spin in an electromagnetic  field. }
Proc. Roy. Soc. London. 1939, Vol. A173, P. 211-232.

\noindent
3.
Dirac P. {\em Relativistic wave  equations.} Proc. Roy. Soc. (London). A. 1936. Vol. 155. P. 447-459.

\noindent
4.
H.J. Bhabha. {\em Relativistic wave equations for elementary particles.}
Rev. Mod. Phys. 1945. Vol. 17. P. 200-216; 1949. vol. 21. P. 451;
{\em On the postulational basis  of the theory  of elementary particles.}  Rev. Mod. Phys. 1949. Vo;. 21.
P.  451-462.

\noindent
5.
Harish-Chandra. Phys. Rev. 1947. Vol. 71. P. 793; Proc. Roy. Soc. (London). 1947.
Vol. 192A. P. 195.

\noindent
6.
H. Umezawa. {\em Quantum field theory.} Amsterdam. 1956.

\noindent
7.
E.M. Corson. {\em Introduction to tensors, spinors and relativistic wave equations.} Blackie. London. 1953.

\noindent
8.
I.M.  Gel'fand, R.A. Minlos, and Z.Ya. Shapiro. {\em Linear representations of the  rotation and Lorentz groups
and their applications.} Pergamon. New York. 1963.

\vspace{1,5mm}
\noindent
9.
M.A. Naimark. {\em Linear representations of the Lorentz group.}  Pergamon. New York, 1964.

\noindent
10.
Krajcik   R., Nieti M.M.  {\em Bhabha firdt-order wave equations.} Phys. Rev. D. 1974. Vol.  10. P.
4049-4062; Phys. Rev. D. 1975. Vol.  11. P. 4042-1458;
Phys. Rev. D. 1975. Vol.  11. P. 4059-1471;
Phys. Rev. D. 1976. Vol.  13. P. 924-941;
Phys. Rev. D. 1976. Vol.  14. P. 418-436;
Phys. Rev. D. 1977. Vol.  15. P. 433-444;
Phys. Rev. D. 1977. Vol.  15. P. 445-452.

\noindent
11.
A.A. Bogush, L.G. Moroz.
{\it Introduction to theory of classical fields.}
Minsk,  1968.  385 pages (in Russian ).

\noindent
12.
F.I. Fedorov. {\it The Lorentz group}.  Moskow. 1989. (in Russian ).

\noindent
13.
V.L. Ginzburg.
{\it On theory of exited spin states of elementary particles.}
JETP. 1943. VOl. 13. P. 33-58 (in Russian).

\noindent
14.
E.S. Fradkin.
{\it On theory of particles with higher spins.}
JETP. 1950. Vol 20. P. 27-38 (in Russian).

\noindent
15.
Petras M.
{\it A note to Bhabha's equation for a particle with maximum spin 3/2.}
Czehc. J. Phys. 1955. Vol. 5. No 3. P. 418-419.

\noindent
16.
V.Ya. Fainberg.
{\it On interaction theory of higher spin particles  with electromagnetic  and meson fields.}
Trudy  FIAN SSSR. 1955. Vol. 6. P. 269-332 (in Russian).

\noindent
17.
I. Ulehla  {\it Anomalous equations for particles with spin 1/2.}
JETP. 1957. Vol. 33. P. 473-477 (in Russian).

\noindent
18.
Formanek J.
{\it On the Ulehla-Petras wave equation.}
Czehc. J. Phys. B.  1955. Vol. 25. No 8. P. 545-553.

\noindent
19.
S.I. Lobko.
{\it On theory particles with variable spin 1/2 -3/2 and with two rest masses.}
Kandid. Dissertation. Minsk,  1965 (in Russian).

\noindent
20.
F.I. Fedorov, V.A. Pletuschov.
{\it Wave equations with repeated representations of the Lorentz group. Integer spin.}
Vesti AN BSSR.  Ser. fiz.-mat. nauk.  1969. No 6. P. 81-88 (in Russian).

\noindent
21.
A.Z. Capri.
{\it Non-uniqueness of the spin 1/2 equation.} Phys. Rev. 1969.  Vol. 178. No 5. P. 1811-1815.

\noindent
22.
A.Z. Capri. {\it First-order  wave equations for half-odd-integral spin.}
Phys. Rev. 1969.  Vol. 178. No 5. P. 2427-2433.

\noindent
23.
A.Z. Capri. {\it First-order  wave equations for multi-mass fermions.}
Nuovo Cimento. B.  1969.  Vol. 64. No 1. P. 151-158.

\noindent
24.
F.I. Fedorov, V.A. Pletuschov. {\it
Wave equations with repeated representations of the Lorentz group. Half-integer spin.}
Vesti AN BSSR. Ser. fiz.-mat. nauk.  1970. No 3. P. 78-83 (in Russian).

\noindent
25.
V.A. Pletuschov, F.I. Fedorov.
{\it Wave equations with repeated representations of the Lorentz group for a spin 0 particle.}
Vesti AN BSSR. Ser. fiz.-mat. nauk.  1970.  No 2. P. 79-85 (in Russian).

\noindent
26.
V.A. Pletuschov, F.I. Fedorov.
{\it Wave equations with repeated representations of the Lorentz group for a spin 1 particle.}
Vesti AN BSSR. Ser. fiz.-mat. nauk. 1970,  No 3. P. 84-92 (in Russian).

\noindent
27.
A. Shamaly, A.Z. Capri. {\it First-order wave equations for integral spin.}
Nuovo Cimento. B. 1971. Vol. 2. No 2. P. 235-253.

\noindent
28.
A.Z. Capri.
{\it Electromagnetic properties of a new spin-1/2 field.}
Progr. Theor. Phys.  1972. Vol. 48. No 4. P. 1364-1374.

\noindent
29.
A. Shamaly, A.Z. Capri.
{\it Unified theories for massive spin 1 fields.}
Can. J, Phys. 1973. Vol. 51. No 14. P. 1467-1470.

\noindent
30.
M.A.K. Khalil.
{\it  Properties of a 20-component spin 1/2 relativistic wave equation. }
Phys. Rev. D.  1977.  Vol. 15. No 6. P. 1532-1539.

\noindent
31.
M.A.K. Khalil.
{\it Barnacle equivalence structure in relativistic wave equation.}
Progr. Theor. Phys.  1978. Vol. 60. No 5. P. 1559-1579.

\noindent
32.
M.A.K. Khalil.
{\it An equivalence of relativistic field equations.}
Nuovo Cimento. A. 1978. Vol. 45. No  3. P. 389-404.

\noindent
33.
M.A.K. Khalil.
{\it Reducible  relativistic wave equations.}
J. Phys. A.: Math. and Gen.  1979. Vol. 12. No  5. P. 649-663.

\noindent
34.
A.A. Bogush, V.V. Kisel.
{\it Equations with repeated representations of the Lorentz group and Pauli interaction.}
Vesti AN BSSR. Ser. fiz.-mat. nauk.   1979.  No 3.  P. 61-65 (in Russian).

\noindent
35.
A.A. Bogush, V.V. Kisel, M.I. Levchuk, L.G. Moroz.
{\it On description of polarizability of scalar particles in the theory of relativistic wave equations.}
In: Covariant methods in  theoretical physics. Elementary particle physics and relativity theory.
Institute of Physics, Academy of sciences of Belarus. Minsk. 1981.  P. 81-90 (in Russian).

\noindent
36.
V.V. Kisel.
{\it Electric polarizability of a spin 1 particle in the theory of relativistic wave  equations.}
Vesti AN BSSR. Ser. fiz.-mat. nauk.  1982. No  3. P. 73-78 (in Russian).

\noindent
37.
A.A. Bogush, V.V. Kisel.
{\it Description of a free particle by different wave equations.}
Doklady AN BSSR.  1984. Vol. 28. No 8. P. 702-705 (in Russian).

\noindent
38.
A.A. Bogush, V.V. Kisel.
{\it Equation for a spin 1/2 particle with anomalous magnetic momentum.}
// Izvestiya VUZov. Fizika. 1984.  No 1. P. 23-27 (in Russian).

\noindent
39.
A.A. Bogush, V.V. Kisel, F.I. Fedorov.
{\it On interpretation of supplementary components of wave functions in presence of electromagnetic
interaction.}
Doklady AN BSSR.  1984. Vol. 277. No 2.  P. 343-346 (in Russian).

\noindent
40.
Lunardi J.T., Pimentel B.M.,  Teixeira R.G.
{\em Duffin-Kemmer-Petiau equation in Rieman\-nian space-times.} -- 14 pages. gr-qc/9909033.

\noindent
41.
Lunardi J. T., Pimentel B.M., Teixeira R. G. and Valverde J.S. 2000.
{\em Remarks on Duffin-Kemmer-Petiau theory and guage invariance.} Phys. Lett. A. 268. 165-73.

\noindent
42.
Fainberg V. Ya. and Pimentel B.M.  {\em Duffin-Kemmer-Petiau  and Klein-Gordon-Fock
equations for electromagnetic, Yang-Mills and external gravitational field interactions:
proof of equivalence.} Phys. Lett. 2000.  A. Vol. 271. P. 16-25.

\noindent
43.
Fainberg V. Ya. and Pimentel B.M. {\em On equivalence Duffin-Kemmer-Petiau
and Klein-Gordon equations.}  Braz. J. Phys. 2000. Vol. 30.  P. 275-81.

\noindent
44.
Fainberg V. Ya. and Pimentel B.M.
{\em Equivalence Duffin-Kemmer-Petiau and Klein-Gordon-Fock equations.} Theor. Math. Phys. 2000.
Vol. 124. P. 1234-49.

\noindent
45.
de Montigny M., Khanna F.C., Santana A. E., Santos E.S. and Vianna J.D.M.
{\em Galilean covariance and the Duffin-Kemmer-Petiau equation.}  J.Phys. A.: Math.
Gen. 2000. Vol. 33. L273-8.

\noindent
46.
Tetrode H. {\em Allgemein relativistishe Quantern theorie des Elektrons.}  Z.
Phys. 1928, Bd 50, S. 336.

\noindent
47.
 Weyl H.
{\em Gravitation and the electron.}
Proc. Nat. Acad. Sci. Amer.  1929.  Vol. 15.  P. 323 -334;
{\em Gravitation and the electron.}
Rice Inst. Pamphlet.  1929.   Vol. 16.   P. 280-295;
{\em Elektron und Gravitation.}  Z. Phys.  1929.   Bd. 56.  S. 330-352;
{\em A remark on the coupling of gravitation and electron.}
 Actas   Akad. Nat. Ciencias  Exactas. Fis. y natur. Lima.  1948.
Vol. 11. P. 1-17.

\noindent
48.
 Fock V., Ivanenko D.
{\em \"{U}ber eine m\"ogliche geometrische Deutung der relativistischen
Quanten\-theorie.}  Z. Phys. 1929. Bd. 54.  S. 798-802;
{\em G\'{e}ometrie   quantique  lin\'{e}aire   et d\'{e}placement parallele.}
 Compt. Rend. Acad. Sci. Paris.  1929.  Vol. 188.  P. 1470-1472.

\noindent
49.
 Fock V.
{\em Geometrisierung der Diracschen Theorie des Elektrons.}
Z. Phys.  1929.  Bd. 57,  S. 261-277.

\noindent
50.
Schouten J.A.
{\em Dirac  equations  in  general  relativity.}
J. Math. and Phys.  1931. Vol. 10.  P. 239-271; P. 272-283.

\noindent
51.
 Schr\"odinger E.
{\em Sur la th\'{e}rie relativiste de l'\'{e}lectron et l'interpr\'{e}tation
de la m\'{e}chanique  quantique.}
Ann. Inst. H. Poincar\'{e}. 1932.  Vol. 2. P. 269-310;
{\em Dirac'sches Elektron  im  Schwerfeld.}
Sitz. Ber. Preuss. Akad. Wiss. Berlin. Phys.-Math Kl.
1932.  S. 105-128.

\noindent
52.
 Einstein  A.,  Mayer  W.
{\em Semivektoren   und   Spinoren.}
Sitz. Ber. Preuss. Akad. Wiss. Berlin. Phys.-Math. Kl.
 1932 . S. 522-550;
{\em Die Diracgleichungen f\"ur Semivektoren.}
Proc. Akad.  Wet. (Amsterdam).  1933.  Bd. 36.  S. 497-516;
{\em Spaltung der Nat\"urlichsten  Feldgleichungen  f\"ur
Semi-Vektoren  in  Spinor-Gleichungen  von  Diracschen Tipus.}
Proc. Akad. Wet. (Amsterdam).  1933.  Bd. 36.  S. 615-619.

\noindent
53.
 Bargmann V.
{\em \"{U}ber der Zusammenhang zwischen  Semivektoren  und
Spinoren  und  die   Reduktion   der   Diracgleichungen   fur Semivektoren.}
Helv. Phys. Acta.  1933.  Bd. 7.  S. 57.

\noindent
54.
 Schouten J.A.
{\em Zur generellen Feldtheorie,  Semi-Vektoren  und Spin-raum.}
Z. Phys.  1933.  Bd. 84.  S. 92-111.

\noindent
55.
Infeld L., van der Waerden B.L.
Die  Wellengleichungen  des
{\em Elektrons in der allgemeinen Relativit\"{a}stheorie.}
Sitz. Ber. Preuss. Akad. Wiss. Berlin. Phys.-Math. Kl.  1933.
Bd. 9.  S. 380-401.

\noindent
56.
 Infeld L.
{\em Dirac's  equation  in  general  relativity  theory.}
Acta Phys. Polon.  1934.  Vol. 3.  P. 1.

\noindent
57.
 Dirac P.A.M.
{em Wave  equations in conformal space.}
Ann. Math. 1936.  Vol. 37.  P. 429-442.

\noindent
58.
 Yamamoto H.
{\em On equations for the Dirac  electron in  general relativity.}
Japan. J. Phys.   1936.  Vol. 11.  P. 35.

\noindent
59.
 Proca A.
{\em Sur la th\'{e}orie ondulatoire des electrons positifs et n\'{e}gatifs.}
J. Phys. et Rad.  1936.  Vol. 7. P. 347.

\noindent
60.
 Benedictus W.
{\em Les \'{e}quations de Dirac dans un  espace \`{a} m\'{e}trique
riemannienne. }
Compt. Rend. Acad. Sci. Paris.  1938.  Vol. 206. P. 1951.

\noindent
61.
 Cartan E.
{\em Le\c{c}ons sur la th\'eorie des  spineurs.}
 Actualit\'es Sci.  1938.

\noindent
62.
 Pauli W.
{\em \"Uber die Invarianz  der Dirac'schen Wellengleichungen gegen\"uber
\"Ahnlichkeit\- stransformationen des Linienelementes im Fall verschwindender
Ruhmasse.}
Helv. Phys. Acta.  1940.  Bd. 13. S. 204-208.

\noindent
63.
 Bade W.L., Jehle H.
{\em An introduction to spinors.}
Rev. Mod. Phys. 1953. Vol. 25. P. 714-728.

\noindent
64.
Brill D.R.,    Wheeler J.A.
{\em Interaction of neutrinos   and gravitational fields.}
Rev. Mod. Phys.  1957.  Vol. 29,  No 3.   P. 465-479.

\noindent
65.
Bergmann  P.G.
{\em Two-component   spinors   in   general relativity.}
Phys. Rev.  1957.  Vol. 107, No 2.  P. 624-629.

\noindent
66.
 Fletcher J.G.
{\em Dirac matrices in Rimannian space.}
Nuovo Cim.  1958.   Vol. 8, No 3.  P. 451-458.

\noindent
67.
 Buchdahl H.A.
{\em On extended  conformal transformations of spinors and spinor
equations.}
Nuovo Cim.   1959.  Vol. 11.  P. 496-506.

\noindent
68.
 Namyslowski J.
{\em The Dirac equation in general relativity in the vierbein formalism.}
Acta Phys. Polon.  1961.  Vol. 20, No 11. P. 927-936;
{\em Symmetrical form of Dirac matrices in general relativity.}
Acta Phys. Polon.  1963.  Vol. 23, No 6.  P. 673-684.

\noindent
69.
Peres A.
{\em Spinor fields in generally covariant theories.}
Nuovo Cim. Suppl.  1962.  Vol. 24, No 2.  P. 389-452.

\noindent
70.
Lichnerowicz  A.
{\em Champ  de  Dirac,  champ du  neutrino  et
transformations $C,P,T$ sur un espace-temps curve.}
Ann. Inst. Henri Poincar\'e. A.  1964, Vol. 1, No 3.  P. 233-290.

\noindent
71.
Ogivetckiy V.I., Polubarinov I.V. {\em On spinors in gravity theory} (in Russian).
JETP.  1965. Vol 48, No 6. P. 1625-1636.

\noindent
72.
Pagels H.
{\em Spin and Gravitation.}
Ann. Phys. (N.Y.)  1965.  Vol. 31, No 1.   P. 64-87.

\noindent
73.
Brill D.R., Cohen J.M.
{\em Cartan frames and general relativistic Dirac equation.}
J. Math. Phys. 1966.  Vol. 7,  No 2. P. 238-245.

\noindent
74.
Cap F., Majerotto W., Raab W., Unteregger P.
{\em Spinor  calculus in Riemannian manifolds.}
Fortschr. der  Physik.   1966.  Bd. 14, No 3.  S. 205-233.

\noindent
75.
Von D. Kramer,  Stephani H.
{\em Bispinorfelder im Riemannschen  Raum.}
Acta Phys. Polonica.  1966.  Vol. 29, No 3.  P. 379-386.

\noindent
76.
Luehr C.P., Rosenbaum M.
{\em Spinor connections in general relativity.}
J. Math. Phys.  1974.  Vol. 15, No 7. P. 1120-1137.

\noindent
77.
Maia M.D.
{\em Conformal  spinor fields in general relativity.|
J. Math. Phys.  1974.  Vol. 15, No 4. P. 420-425.

\noindent
78.
Brum\^a C. {\em A mode of constructing the  Dirac matrices in gravitational
field.}  Rev. Roum. Phys.  1986.  Tome 31,  No 8.  P. 753-763;
{\em  On Dirac matrices in gravitational field.}
 Rev. Roum. Phys.  1987.  Tome 32,  No 4.  P. 375-382.

\noindent
79.
Fariborz A.H., McKeon D.G.C. {\em  Spinors in Weyl geometry.}
Class. Quant. Grav.  1997.  Vol. 14.  P. 2517-2525;  hep-th/9607056.

\noindent
80.
Manoelito  M. de Souza. {\em  The Lorentz Dirac equation and the structure of
spacetime.}   Braz. J. Phys.  1998.  Vol. 28.  P. 250-256;
hep-th/9505169.

\end{document}